\documentclass[fleqn,usenatbib]{mnras}

\usepackage[T1]{fontenc}

\DeclareRobustCommand{\VAN}[3]{#2}
\let\VANthebibliography\thebibliography
\def\thebibliography{\DeclareRobustCommand{\VAN}[3]{##3}\VANthebibliography}

\usepackage{graphicx}	
\usepackage{amsmath}	
\usepackage{amssymb}	

\def\Lya{Ly$\alpha$~}

\def\21cm{$21\rm\,cm$}

\def\HI{\hbox{H$\,\rm \scriptstyle I$}}
\def\HII{\hbox{H$\,\rm \scriptstyle II$}} 
\def\HeI{\hbox{He$\,\rm \scriptstyle I$}}
\def\HeII{\hbox{He$\,\rm \scriptstyle II$}}
\def\HeIII{\hbox{He$\,\rm \scriptstyle III$}} 
\def\OI{\hbox{O$\,\rm \scriptstyle I$}} 
\def\MgII{\hbox{Mg$\,\rm \scriptstyle II$}}

\usepackage{newtxtext,newtxmath}

\title[Strong \boldmath{21\,cm} forest absorbers]{The detectability of strong \boldmath{21} centimetre forest absorbers  from the diffuse intergalactic medium in late reionisation models}

\author[T. Šoltinský et al.]{Tomáš Šoltinský$^{1}$\thanks{E-mail: tomas.soltinsky@nottingham.ac.uk},
James S. Bolton$^{1}$,
Nina Hatch$^{1}$,
Martin G. Haehnelt$^{2}$, \newauthor
Laura C. Keating$^{3}$,
Girish Kulkarni$^{4}$,
Ewald Puchwein$^{3}$,
Jonathan Chardin$^{5}$ \newauthor 
\& Dominique Aubert$^{5}$
\\
$^{1}$School of Physics and Astronomy, University of Nottingham, University Park, Nottingham, NG7 2RD, UK\\
$^{2}$Kavli Institute for Cosmology and Institute of Astronomy, Madingley Road, Cambridge, CB3 0HA, UK\\
$^{3}$Leibniz-Institut f\"ur Astrophysik Potsdam, An der Sternwarte 16, 14482 Potsdam, Germany\\
$^{4}$Tata Institute of Fundamental Research, Homi Bhabha Road, Mumbai 400005, India\\
$^{5}$Observatoire Astronomique de Strasbourg, 11 rue de l’Universite, 67000 Strasbourg, France\\}
\date{\today}

\pubyear{2020}

\begin{document}
\label{firstpage}
\pagerange{\pageref{firstpage}--\pageref{lastpage}}
\maketitle

\begin{abstract}
A late end to reionisation at redshift $z\simeq 5.3$ is consistent with observed spatial variations in the \Lya forest transmission and the deficit of \Lya emitting galaxies around extended \Lya absorption troughs at $z=5.5$.  In this model, large islands of neutral hydrogen should persist in the diffuse intergalactic medium (IGM) until $z\simeq 6$.  We use a novel, hybrid approach that combines high resolution cosmological hydrodynamical simulations with radiative transfer to predict the incidence of strong \21cm forest absorbers with optical depths $\tau_{21}>10^{-2}$ from the diffuse IGM in these late reionisation models.  We include the effect of redshift space distortions on the simulated \21cm forest spectra, and treat the highly uncertain heating of the pre-reionisation IGM by soft X-rays as a free parameter.  For a model with only modest IGM pre-heating, such that average gas kinetic temperatures in the diffuse IGM remain below $T_{\rm K}\simeq 10^{2} \rm\, K$, we find that strong \21cm forest absorption lines should persist until $z=6$.  For a sample of $\sim 10$ sufficiently radio loud background sources, a null-detection of \21cm forest absorbers at $z\simeq 6$ with SKA1-low or possibly LOFAR should provide an informative lower limit on the still largely unconstrained soft X-ray background at high redshift and the temperature of the pre-reionisation IGM.
\end{abstract}



\section{Introduction}

At present, the premier technique for examining the small-scale structure of intergalactic neutral hydrogen approaching the reionisation era is Lyman series absorption in the spectra of luminous quasars \citep{Becker_2015,Eilers_2017,Bosman_2018,Yang_2020_z63}. However, it is challenging to probe the intergalactic medium (IGM) much beyond redshift $z\simeq 6$ with this approach. The large cross-section for \Lya scattering means the IGM becomes opaque to \Lya photons at neutral hydrogen fractions as low as $x_{\rm HI}\simeq 10^{-4}$.   An alternative transition that overcomes this limitation is the hyperfine \21cm line, which has a cross-section that is a factor $\sim 10^{7}$ smaller than the \Lya transition\footnote{Low ionisation metal lines such as \OI~ \citep{Oh_2002,Keating_2014} and \MgII~\citep{Hennawi_2020} can also be used to trace neutral intergalactic gas, although the uncertain metallicity of the high redshift IGM further complicates their interpretation.}. If radio bright sources such as high redshift quasars \citep{Banados_2021} or gamma-ray bursts \citep[e.g.][]{Ioka_2005,Ciardi_2015_GRB} can be identified during the reionisation era, the intervening neutral IGM may be observed as a \21cm forest of absorption lines in their spectra.  This can be achieved either through the direct identification of individual absorption features \citep{Carilli_2002,Furlanetto_2002,Meiksin_2011,Xu_2011,Ciardi_2013,Semelin_2016,VillanuevaDomingo2021}, or by the statistical detection of the average \21cm forest absorption  \citep{Mack_2012,Ewall_Wice_2014,Thyagaragan_2020}.   This approach is highly complementary to proposed tomographic studies of the redshifted \21cm line and measurements of the \21cm power spectrum during reionisation \citep[e.g.][]{Mertens_2020,Trott_2020}, as it is subject to a different set of systematic uncertainties \citep{Furlanetto_Oh_2006,Pritchard_2012}.

However, any detection of the \21cm forest relies on the identification of sufficient numbers of radio-loud sources and the existence of cold, neutral gas in the IGM at $z\gtrsim 6$.  While neither of these criteria are guaranteed, the prospects for both have improved somewhat in the last few years.  Approximately $\sim 10$ radio-loud active galactic nuclei are now known at $5.5<z<6.5$ 
\citep[e.g.][]{Banados_2018,Banados_2021,Liu_2020,Ighina_2021}, including the $z=6.1$ blazar PSO J0309+27 with a flux density $S_{147\rm\,MHz}=64.2\pm 6.2\rm\,mJy$ \citep{Belladitta_2020}.  The Low Frequency Array (LOFAR) Two-metre Sky Survey  \citep[LoTSS,][]{Shimwell_2017,Kondapally_2020}, the Giant Metrewave Radio Telescope (GMRT) all sky radio survey at $150\rm\,MHz$ \citep{Intema_2017}, and the Galactic and Extragalactic All-sky Murchison Widefield Array survey \citep[GLEAM,][]{Wayth_2015} are also projected to detect hundreds of bright $z>6$ radio sources, particularly if coupled with large spectroscopic follow-up programmes such as the William Herschel Telescope Enhanced Area Velocity Explorer (WEAVE)-LOFAR survey \citep{Smith_2016_WEAVE}.

Furthermore, there is now growing evidence that reionisation ended rather late, and possibly even extended to redshifts as late as $z\simeq 5.3$ \citep{Kulkarni_2019,Nasir_2020,Qin_2021}.  This picture is motivated by the large spatial fluctuations observed in the \Lya forest transmission at $z\simeq 5.5$ \citep{Becker_2015,Eilers_2018}.  A late end to reionisation is also consistent with the electron scattering optical depth inferred from the cosmic microwave background \citep{Pagano_2020}, the observed deficit of \Lya emitting galaxies around extended \Lya absorption troughs \citep{Becker_2018,Kashino_2020,Keating_2020}, the clustering of \Lya emitters \citep{Weinberger_2019}, the thermal widths of \Lya forest transmission spikes at $z>5$ \citep{Gaikwad_2020}, and the mean free path of ionising photons at $z=6$ \citep{Becker_2021}.  If this interpretation proves to be correct \citep[but see][for alternative explanations]{DAloisio_2015,Davies_2016,Chardin_2017,Meiksin_2020}, then there should still be large islands of neutral hydrogen in the IGM as late as $z=6$ \citep[e.g.][]{Lidz_2007,Mesinger_2010}.  If this neutral gas has not already been heated to kinetic temperatures $T_{\rm K}\gtrsim 10^{3}\rm\, K$ by the soft X-ray background, then it may be possible to detect \21cm absorbers in the pre-reionisation IGM at $z\simeq 6$.  Alternatively, a null-detection could provide a useful lower limit on the soft X-ray background at high redshift.

The goal of this work is to investigate this possibility further.  We use a set of high resolution hydrodynamical cosmological simulations drawn from the Sherwood-Relics simulation programme (Puchwein et al. in prep).  Using a novel hybrid approach, these are combined with the ATON radiative transfer code \citep{Aubert_2008} to model the small-scale structure of the IGM.   Following \citet{Kulkarni_2019}, we consider a model with late reionisation ending at $z=5.3$, and contrast this with  a simulation that has an earlier end to reionisation at $z=6.7$.  We pay particular attention to some of the common assumptions adopted in previous models of the \21cm forest that affect the absorption signature on small scales.  This includes the treatment of gas peculiar motions and thermal broadening, the coupling of the spin temperature to the \Lya background, and the effect of pressure (or Jeans) smoothing on the IGM.  Our approach is therefore closest to the earlier work by \citet{Semelin_2016}, although we do not follow spatial variations in the X-ray and \Lya backgrounds.  Instead, we attempt to explore a broader range of parameter space for spatially uniform X-ray heating using hydrodynamical simulations that use several different reionisation histories and have an improved mass resolution (by a factor $\sim 27$).  Note, however, that even the high resolution cosmological simulations considered here will still only capture the \21cm absorption that arises from the diffuse IGM.  We therefore do not model the (uncertain amount of) absorption from neutral gas in haloes below the atomic cooling threshold, or from the cold interstellar medium in much rarer, more massive haloes that host high redshift galaxies \citep[see e.g.][]{Furlanetto_2002,Meiksin_2011}. 

This paper is structured as follows.  We start by describing our numerical model of the IGM in Section~\ref{sec:IGM_model}, and our calculation of the \21cm optical depths in Section~\ref{sec:21cm_modeling}. We examine how different modelling assumptions affect the observability of strong  \21cm forest absorbers in Section~\ref{sec:tau_ave} and~\ref{sec:diffdist}, and estimate how a null-detection of strong \21cm absorbers at redshift $z\simeq 6$ with LOFAR or the Square Kilometre Array (SKA) could constrain the high redshift soft X-ray background in Section~\ref{sec:observability}. Finally, we conclude in Section~\ref{sec:conclusion}. The appendix contains further technical details regarding our methodology and modelling assumptions.


\section{Numerical model for the 21 cm forest}\label{sec:IGM_model}
\subsection{Hydrodynamical simulations with radiative transfer}\label{sec:simulation}

We model the $21\rm\, cm$ forest during inhomogeneous reionisation using a sub-set of the high resolution cosmological hydrodynamical simulations drawn from the Sherwood-Relics simulation programme \citep[see][for an initial application of these models]{Gaikwad_2020}.  The Sherwood-Relics simulations were performed with a modified version of the \textsc{P-GADGET-3} code -- which is itself an updated version of the \textsc{GADGET-2} code described in \citep{Springel_2005} -- and uses the same initial conditions as the earlier Sherwood simulation suite \citep{Bolton_2017}.  In this work we adopt a flat $\Lambda$CDM cosmology with $\Omega_{\Lambda}=0.692$, $\Omega_{\rm m}=0.308$, $\Omega_{\rm b}=0.0482$, $\sigma_8=0.829$, $n_{\rm s} =0.961$, $h=0.678$, consistent with \citet{planck2014}, and a primordial helium fraction by mass of $Y_{\rm p}=0.24$ \citep{Hsyu_2020}.   

The simulations have a volume $(40h^{-1}$cMpc$)^3$ and track $2\times2048^3$ dark matter and gas particles.  This yields a dark matter particle mass of  $7.9\times10^5 M_{\odot}$ and resolves dark matter haloes with masses greater than $\sim 2.5\times10^7 M_{\odot}$.  This high mass resolution is necessary for capturing the small-scale intergalactic structure probed by the \21cm forest \citep[cf.][]{Semelin_2016}.   We furthermore adopt a simple but computationally efficient scheme for converting high density gas into collisionless particles that robustly predicts the properties of the IGM.  If a gas particle has an overdensity $\Delta=1+\delta>1000$ and kinetic temperature $T_{\rm K}<10^5 \rm\, K$, it is converted into a collisionless star particle \citep{Viel_2004}. We have verified this simplified approach is sufficient for modelling the 21cm forest in the diffuse IGM by direct comparison to a full sub-grid star formation model (see Appendix~\ref{sec:starformation} for further details). The main effect of this approximation is the removal of dense gas from haloes, which slightly reduces the number of strong $21\rm\, cm$ absorbers in models with no X-ray heating.

\begin{table}
	\centering
	\caption{Hydrodynamical simulations used in this work. From left to right, the columns give the name of the simulation, the nature of the reionisation model, the redshift when the IGM is fully reionised, and the prescription for converting dense gas into collisionless star particles, which follows either \citet{Viel_2004} or \citet{Puchwein_2013}. The first three simulations are part of the Sherwood-Relics suite.  The final two models are optically thin simulations with rapid reionisation at $z\simeq 15$, and are described in Appendix~\ref{sec:starformation}.  All models have a volume of $(40h^{-1}$cMpc$)^3$ and include $2\times2048^3$ dark matter and gas particles.}
	\label{tab:sims}
	\begin{tabular}{lccc}
		\hline
		Name & Reionisation & $z_{\rm r} $ & Star formation\\
		\hline
		zr53         & Hybrid RT/hydro           & 5.3        & VHS04\\
		zr53-homog   & Homogeneous, matches zr53 & 5.3        & VHS04\\
		zr67         & Hybrid RT/hydro           & 6.7        & VHS04\\
		Q\Lya        & Rapid, optically thin     & $\sim 15$  & VHS04\\
		PS13         & Rapid, optically thin     & $\sim 15$  & PS13\\
		\hline
	\end{tabular}
\end{table}

In order to include the effect of inhomogeneous reionisation by UV photons on the IGM, the Sherwood-Relics simulations are combined with the moment-based, M1-closure radiative transfer code ATON \citep{Aubert_2008}.  We adopt a novel hybrid approach that captures the small-scale hydrodynamical response of the gas in the simulations to patchy heating during reionisation \citep[see also][for a related approach]{Onorbe_2019}.  Our hybrid RT/hydrodynamical simulations use inputs in the form of 3D maps of the reionisation redshift and \HI\ photo-ionisation rate, produced by ATON simulations performed on the \textsc{P-GADGET-3} outputs in post-processing. These maps are then fed back into a re-run of the \textsc{P-GADGET-3} model, where they are called within a non-equilibrium thermo-chemistry solver \citep{Puchwein_2015}.  Following \citet{Kulkarni_2019}, the ionising sources in the ATON simulations have luminosities proportional to the halo mass with a redshift-dependent normalization, and the mean energy of ionising photons is assumed to be $18.6\rm\, eV$.  Further details can be found in \citet{Gaikwad_2020} and Puchwein et al. (in prep).   The main advantage of this approach is that since the post-processing step using the ATON radiative transfer simulations is computationally cheap compared to the hydrodynamical simulations, we may empirically calibrate the source model to yield a reionisation history that is consistent with a wide range of observational constraints. This avoids many of the uncertainties associated with direct hydrodynamical modelling of the source population.

Note, however, the relatively small volume of our simulations means the patchy structure of reionisation will not be fully captured on scales larger than the box size.  This will lead to smaller neutral islands and an earlier percolation of ionised regions relative to simulations performed in a larger volume \citep{Iliev_2014,Kaur_2020}.  We therefore adjust the ionising emissivity in the models by hand to achieve a given reionisation history; this scaling is equivalent to varying the escape fraction of ionising photons.  In addition,  while our ATON simulations self-consistently follow the propagation of ionising photons using a $2048^{3}$ Cartesian grid, self-shielded regions below the size of the grid cells ($\lesssim 20h^{-1}\rm\,ckpc$) will not be resolved.  We attempt to partially correct for this by implementing a correction for the self-shielding of dense gas in all our simulations in post-processing, using the results of \citet{Chardin_2018}.  We find, however, that this correction makes almost no difference to our final results, as the majority of the strong \21cm absorbers in our simulations arise from the diffuse IGM.

\begin{figure}
    \begin{minipage}{\columnwidth}
 	  \centering
 	  \includegraphics[width=\linewidth]{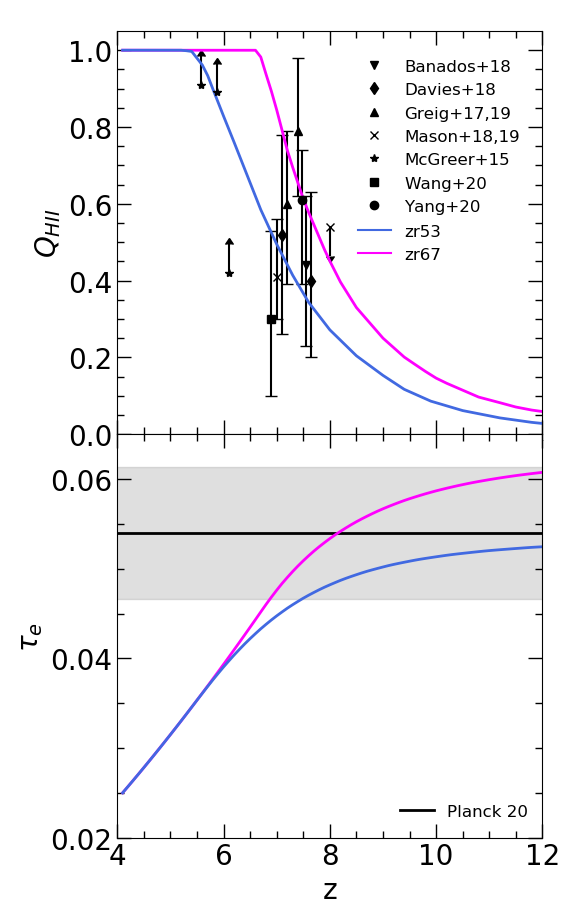}
	\end{minipage}
	\vspace{-0.5cm}
    \caption{\textit{Top}: The filling factor of ionised gas in the zr53 (blue solid curve) and zr67 (fuchsia solid curve) simulations, compared to observational constraints from dark gaps in the \Lya forest \citep{McGreer_2015}, the damping wing in high redshift quasar spectra \citep{Banados_2018_dampwing,Davies_2018,Greig_2017,Greig_2019,Wang_2020,Yang_2020_z75} and \Lya emitting galaxies \citep{Mason_2018,Mason_2019}. For clarity of presentation a small offset has been added to the redshifts of some of the data points. \textit{Bottom}: The Thomson scattering optical depth to cosmic microwave background photons. The black line with the shaded region shows the \citet{Planck_2020} measurement.}
    \label{fig:intro_sim}
\end{figure}

We consider two different reionisation histories in this work, where the IGM becomes fully ionised at either $z=5.3$ (model zr53) or $z=6.7$ (model zr67).  The filling fraction of ionised gas and the Thomson scattering optical depth predicted by these models are displayed in Fig.~\ref{fig:intro_sim}.   Both of these models are consistent with current observational constraints on the timing of reionisation.  As already discussed, the reionisation model that ends at $z=5.3$ is furthermore consistent with the large fluctuations in the \Lya forest transmission observed at $z\simeq 5.5$ \citep{Becker_2015,Kulkarni_2019,Keating_2020}.  Finally, we also use a simulation (zr53-homog) that has been constructed to give exactly the same globally averaged reionisation history as the zr53 model, but using a spatially uniform ionising background.  A comparison between the zr53 and zr53-homog simulations therefore allows us to estimate the uncertain effect that pressure smoothing (from e.g. UV photo-heating) may have on the gas in the pre-reionisation IGM (see Section~\ref{sec:diffdist}).   All the simulations used in this work are listed in Table~\ref{tab:sims}, where the final two models listed are used in Appendix~\ref{sec:starformation} only.

\subsection{Heating of neutral gas by the X-ray and \texorpdfstring{Ly$\alpha$}{} backgrounds}\label{sec:Xray_back}

Absorption features in the $21\rm\,cm$ forest arise from neutral hydrogen in the IGM.  In addition to modelling the inhomogeneous reionisation of the IGM by UV photons, we must therefore also consider the temperature and ionisation state of gas that is optically thick to Lyman continuum photons.  This heating is attributable to adiabatic compression and shocks -- which are already included within our hydrodynamical simulations -- and the X-ray and (to a lesser extent) \Lya radiation backgrounds at high redshift \citep{Ciardi_2010}, which are not.   Hence, we now describe the procedure we use to include spatially uniform X-ray and \Lya heating in our simulations, by recalculating the density dependent temperature and ionisation state of the neutral gas in our hybrid simulations in post-processing.    Further details on the model we use are also provided in Appendix~\ref{sec:IGM_heating}.  

As we do not directly model the star formation rate in our simulations, rather than using a detailed model for the number and spectral energy distribution of X-ray sources at high redshift, for simplicity and ease of comparison to the existing literature we instead follow the approach introduced by \citet{Furlanetto_2006b} for parameterising the comoving X-ray background emissivity. This uses the observed relationship between the star formation rate, $\rm SFR$, and hard X-ray band luminosity ($2$--$10\rm\,keV$) for star-forming galaxies at $z=0$ \citep{Gilfanov_2004,Lehmer_2016}. \citet{Furlanetto_2006b} adopt the normalisation
\begin{equation}
L_{\rm X}=3.4\times 10^{40}\rm\,erg\,s^{-1} \, \mathnormal{f_{\rm X}} \left(\frac{SFR}{1\,M_{\odot}\rm\,yr^{-1}}\right), \label{eq:LX} 
\end{equation}
\noindent
for the total X-ray luminosity at photon energies $>0.2\rm\,keV$, assuming a power-law spectral index $\alpha_{\rm X}=1.5$.  The X-ray efficiency, $f_{\rm X}$, parameterises the large uncertainty in the extrapolation of Eq.~(\ref{eq:LX}) toward higher redshift.  Using the conversion $\epsilon_{\rm X, 0.2\rm\,keV}=L_{\rm X}(\alpha_{\rm
  X}-1)/\nu_{0.2 \rm keV}$, the corresponding comoving X-ray emissivity is
\begin{align} \epsilon_{\rm X, \nu}(z) =~& 3.5\times 10^{21}f_{\rm X} \rm\,erg\,s^{-1}\,Hz^{-1}\,cMpc^{-3} \nonumber \\  &\times \left(\frac{\nu}{\nu_{\rm 0.2\rm\,keV}}\right)^{-\alpha_{\rm X}}\left(\frac{\rho_{\rm SFR}(z)}{10^{-2} \,M_{\odot}\rm\,yr^{-1}\,cMpc^{-3}}\right). \label{eq:epsilonX} \end{align}
\noindent
We assume a power-law spectrum with $\alpha_{\rm X}=1.5$, and use the fit
to the observed comoving star formation rate density from
\citet{Puchwein_2019} (their eq. 21), where
\begin{equation} \rho_{\rm SFR}(z) = 0.01{\rm \,M_{\odot}\,yr^{-1}\,cMpc^{-3}} \frac{(1+z)^{2.7}}{1+[(1+z)/3.0]^{5.35}}. \end{equation}
\noindent
We assume that $\rho_{\rm SFR}=0$ at redshifts $z>z_{\star}=14$,
and have verified that adopting $z_{\star}>14$ does not change our
predictions for $21\rm\,cm$ absorption at $z\leq 12$.

The UV background emissivity at the \Lya wavelength from stars in our model is instead given by
\begin{align} \epsilon_{\alpha}(z) =~& 8.7\times 10^{25} f_{\alpha} \rm\,erg\,s^{-1}\,Hz^{-1}\,cMpc^{-3} \nonumber \\ &\times \left(\frac{\rho_{\rm SFR}(z)}{10^{-2}\,M_{\odot}\rm\,yr^{-1}\,cMpc^{-3}}\right), \label{eq:Lyaemis} \end{align}
\noindent
where we have used the conversion between SFR and UV luminosity at
$1500$ \AA\ from \citet{Madau_Dickinson_2014} and assumed a flat
spectrum $(\epsilon_{\nu}\propto \nu^{0})$ in the UV, where the \Lya efficiency $f_{\alpha}$ parameterises the uncertain amplitude.  We adopt $f_{\alpha}=1$ as the fiducial value in this work, but note that this parameter is uncertain and the \Lya emissivity should furthermore vary spatially \citep[see e.g. Fig. 4 in][]{Semelin_2016}.  For illustrative purposes we therefore also show some results for the much smaller value of $f_{\alpha}=0.01$ (but note that in practise $f_{\alpha}$ and the reionisation history are not fully decoupled).   The primary effect of increasing (decreasing) the \Lya efficiency is to produce a tighter (weaker) coupling of the
\HI\ spin and kinetic temperatures. A smaller value of $f_{\alpha}$ may be more appropriate for absorbers that are distant from the sources of \Lya background photons. Instead of a flat UV spectrum we also considered the power-law population II and III spectra used by \citet{Pritchard_Furlanetto_2006},
but the strength of the \Lya coupling in our model is not very sensitive
to this choice at the redshifts we consider.

\begin{figure}
\begin{center}
  \includegraphics[width=0.47\textwidth]{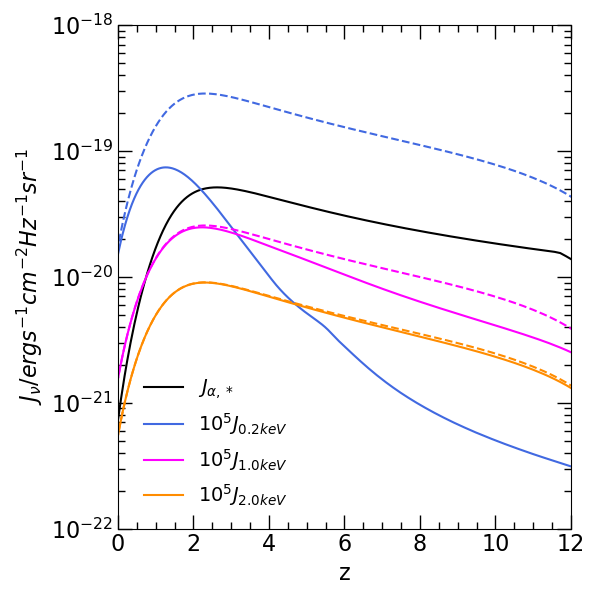}
  \vspace{-0.5cm}
  \caption{The redshift evolution of the specific intensity of the
    \Lya background from stars for a \Lya efficiency $f_{\alpha}=1$ (solid black curve) and the specific intensity of the X-ray
    background for photon energies $0.2\rm\,keV$ (blue curve),
    $1\rm\,keV$ (fuchsia curve) and $2\rm\,keV$ (orange curve), assuming an X-ray efficiency of $f_{\rm X}=1$.  The X-ray specific intensities have been multiplied by a factor of $10^{5}$ for presentation purposes. For comparison, the dashed curves show the X-ray specific intensities evaluated in the optically thin limit. }
  \label{fig:Jnu_evol}
\end{center}
\end{figure}

\begin{figure}
    \begin{minipage}{\columnwidth}
 	  \centering
 	  \includegraphics[width=\linewidth]{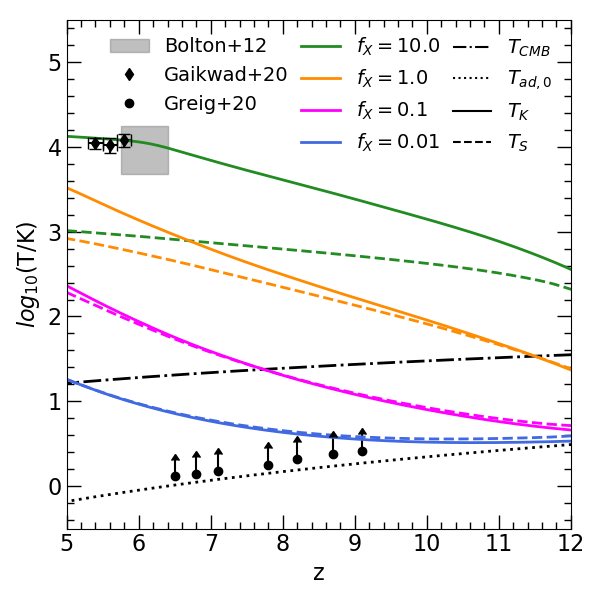}
	\end{minipage}
	 \vspace{-0.5cm}
    \caption{The redshift evolution of the gas kinetic temperature, $T_{\rm K}$, (solid curves) and spin temperature, $T_{\rm S}$, (dashed curves) at mean density following X-ray background heating by photons with $E=0.2$--$30\rm\, keV$.  The different coloured curves correspond to efficiency parameter $f_{\rm X}=10$ (green curves), $1$ (orange curves), $0.1$ (fuchsia curves) and $0.01$ (blue curves). For comparison, the CMB temperature, $T_{\rm CMB}=2.73\rm\, K(1+\mathnormal{z})$ corresponds to the dot-dashed curve, and the kinetic temperature for adiabatic heating and cooling only, $T_{\rm ad,0}=2.73\rm\, K(1+z)^2/(1+\mathnormal{z_{\rm dec}})$, is shown by the dotted curve.   We assume the gas thermally decouples from the CMB at $z_{\rm dec}=147.8$ \citep{Furlanetto_Oh_2006}. The filled diamonds and grey shading correspond to the gas kinetic temperature measurements from Ly$\alpha$ transmission spikes in quasar spectra \citep{Gaikwad_2020} and \Lya absorption lines in quasar proximity zones \citep{Bolton_2012}, respectively. The filled circles show the model dependent lower limits on the \HI\ spin temperature obtained from LOFAR \citep{Greig_2020_LOFAR} and MWA \citep{Greig_2020_MWA}.} 
    \label{fig:Tevol}
\end{figure}

With these emissivities in hand, we may evaluate the solution to the cosmological radiative transfer equation (see Eq.~(\ref{eq:Jnu}) in Appendix~\ref{sec:IGM_heating}) to obtain the X-ray specific intensity at photon energies $0.2$--$30\rm\,keV$ \citep{Pritchard_2012}.  Similarly, we obtain the specific intensity of the \Lya background by evaluating Eq.~(\ref{eq:Jnu_alpha}), following \citet{Pritchard_Furlanetto_2006}.  Fig.~\ref{fig:Jnu_evol} shows the redshift evolution of the specific intensity of the \Lya background from stellar emission, $J_{\alpha,\star}(z)$, and the specific intensity of the X-ray
background at three different energies, $0.2\rm\,keV$, $1\rm\,keV$ and
$2\rm\,keV$.  The dashed curves show the X-ray specific intensities in
the optically thin limit, i.e. when the optical depth of the intervening IGM to X-ray photons is set to zero in Eq.~(\ref{eq:Jnu}).  Note that $J_{2.0\rm\,keV}$ remains almost unchanged in the optically thin limit, implying the IGM is transparent to photons emitted with energies $\geq 2\rm\,keV$ at $z\la 10$ \citep[cf.][]{McQuinn_2012}.

The unresolved soft X-ray background at $z=0$ places an upper limit on the contribution of high redshift
sources to the hard X-ray background, since these photons may redshift
without significant absorption to $z=0$ \citep{Dijkstra_2004,McQuinn_2012}.  When assuming $f_{\rm X}=1.8$, integrating our model specific intensity in the
soft X-ray band ($0.5$-$2\rm\,keV$) at $z=0$ yields $J_{0.5-2\rm
  keV}=2.9\times 10^{-12}\rm\,erg\,s^{-1}cm^{-2}\,deg^{-2}$.  This
value is consistent with the unresolved soft X-ray background obtained from Chandra observations of the COSMOS legacy field, $J_{0.5-2\rm keV}=2.9\pm 0.16 \times
10^{-12}\rm\,erg\,s^{-1}cm^{-2}\,deg^{-2}$
\citep{Cappelluti_2017}.  Note, however, the $z=0$ soft X-ray background does not provide a direct constraint on
the very uncertain soft X-ray background at high redshift \citep[see e.g.][]{Dijkstra_2012,Fialkov_2017}.  Recently, \citet{Greig_2020_MWA} have presented the first weak, model dependent lower limits on the soft X-ray background emissivity at $6.5\leq z \leq 8.7$ using the Murchison Widefield Array (MWA) upper limits on the \21cm power spectrum \citep{Trott_2020}, where $\epsilon_{\rm X,0.5-2\rm\,keV}\gtrsim 10^{34.5} \rm\rm erg\,s^{-1}\,cMpc^{-3}$. For comparison, for an X-ray efficiency of $f_{\rm X}=0.01$, our X-ray background model gives $\epsilon_{\rm X,0.5-2\rm\,keV}=10^{36.0} \rm\, erg\,s^{-1}\,cMpc^{-3}$ at $z=8.1$, which is well above the \citet{Greig_2020_MWA} lower limit.

\begin{figure*}
    \begin{minipage}{2\columnwidth}
 	  \centering
 	  \includegraphics[width=\linewidth]{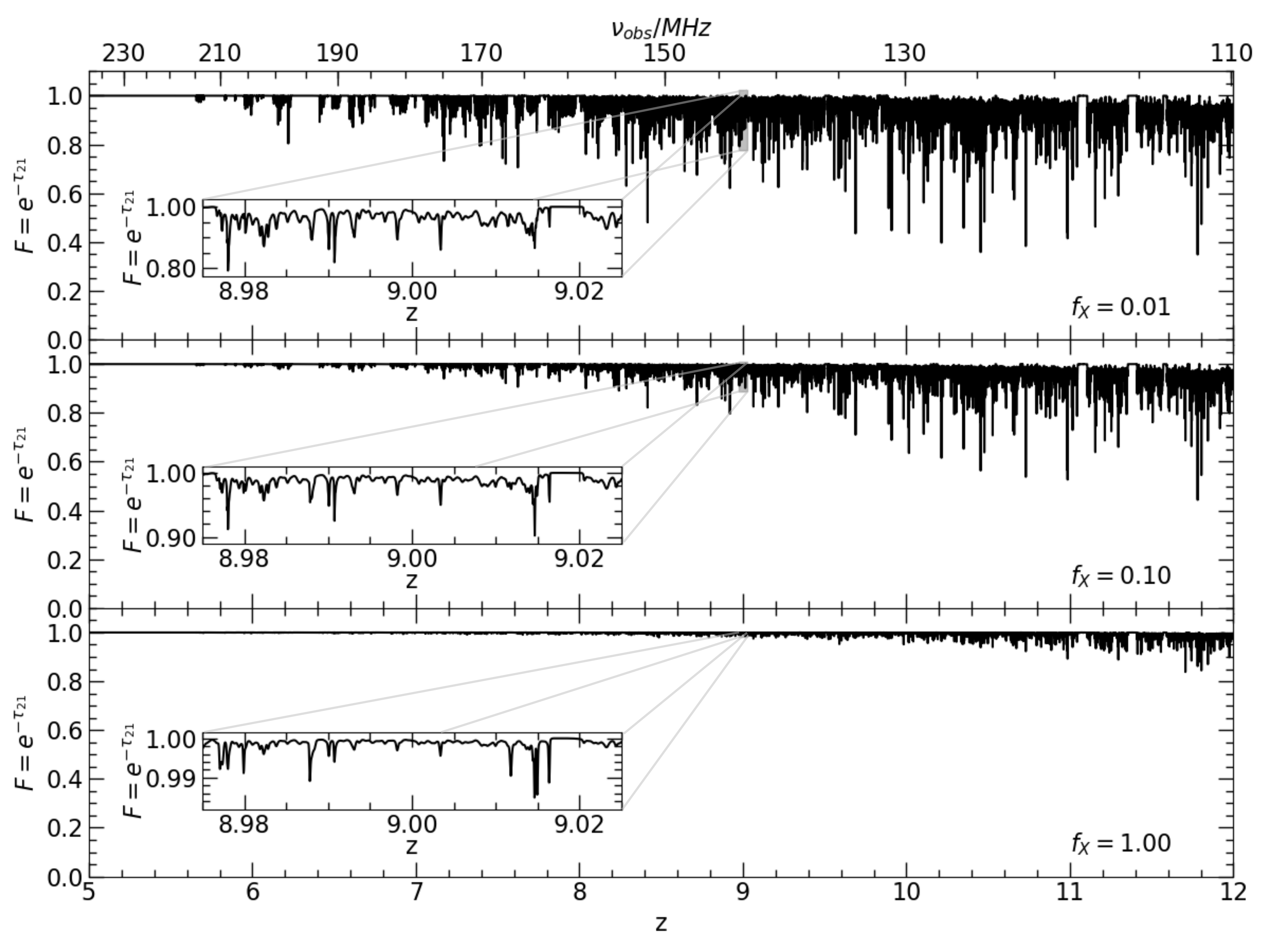}
	\end{minipage}
	\vspace{-0.3cm}
    \caption{The redshift evolution of the \21cm forest transmission, $F=e^{-\tau_{21}}$, in the zr53 simulation for a \Lya efficiency $f_{\alpha}=1$ and an X-ray efficiency of $f_{\rm X}=0.01$ (top), $0.1$ (middle) and $1$ (bottom). The inset displays a zoom-in on part of the \21cm forest at redshift $z\sim9$ -- note the different scales on the vertical axes of the inset.  The incidence of gaps in the \21cm forest, which are associated with large regions of ionised gas, increases toward lower redshift, and become particularly apparent in the $f_{\rm X}=0.01$ model at redshift $z<7$. No instrumental features have been added to the spectra.} 
    \label{fig:long_spectrum}
\end{figure*}

\begin{figure}
    \begin{minipage}{\columnwidth}
 	  \centering
 	  \includegraphics[width=\linewidth]{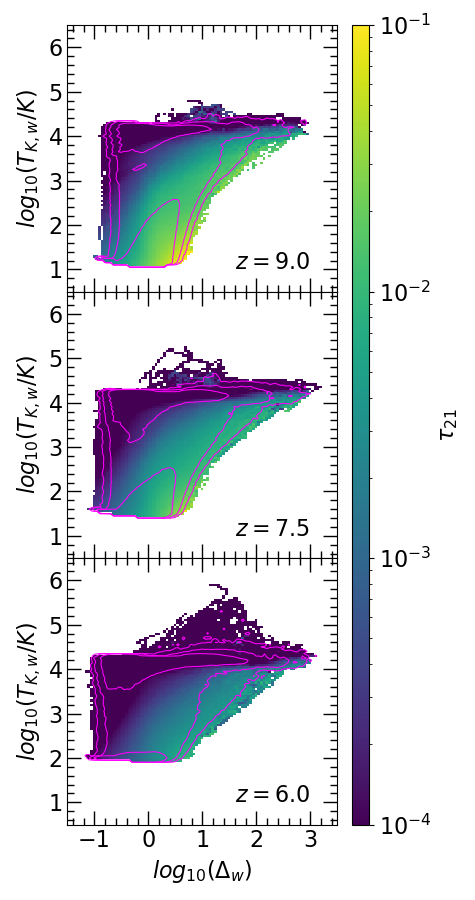}
	\end{minipage}
	\vspace{-0.5cm}
    \caption{The optical depth weighted temperature-density plane for gas in the zr53 simulation at redshift $z=9$ (top), 7.5 (middle) and 6 (bottom), for an X-ray efficiency $f_{\rm X}=0.1$ and \Lya efficiency $f_{\alpha}=1$.  The colour scale shows the average \21cm optical depth at each point in the plane.  The number density of points increases by $1\rm\, dex$ for each contour level.} 
    \label{fig:strong_absorbers}
\end{figure}


Given the specific intensities of the X-ray and \Lya radiation backgrounds, we next compute the thermal evolution of the IGM that remains optically thick to UV photons, but is heated by X-ray and \Lya backgrounds that are assumed to be spatially uniform on the scale of our simulated volume.\footnote{The mean free path to X-ray photons is $\lambda_{\rm X} = 5\rm\,cMpc \,x_{\rm HI}^{-1}(1+\delta)^{-1}(E/0.2\rm\,keV)^{3}[(1+z)/10]^{-2}$.  Fluctuations in the temperature of soft X-ray heated gas on $\sim10\rm\,cMpc$ scales are thus expected \citep{Pritchard_Furlanetto_2007,Ross_2017,Eide_2018}.  These fluctuations would not, however, be adequately captured in our small simulation volume.}   We follow the procedure described in Appendix~\ref{sec:IGM_heating} for this purpose.  Fig.~\ref{fig:Tevol} displays the temperature evolution of a gas parcel at mean density for four different values of the X-ray efficiency parameter $f_{\rm X}$.    An approximate lower limit on $f_{\rm X}$ is provided by the recent constraints on the spin temperature from upper limits on the $21\rm\,cm$ power spectrum at $z\simeq 9.1$ obtained with LOFAR \citep{Mertens_2020}, and at $z=6.5-8.7$ from MWA \citep{Trott_2020}. These data disfavour very cold reionisation models with no X-ray heating \citep{Greig_2020_LOFAR,Mondal_2020,Ghara_2020,Greig_2020_MWA}.  An approximate upper limit on $f_{\rm X}$ at $z>6$ is provided by \Lya absorption measurements of the kinetic temperature at $z \simeq 5$--$6$, after the IGM has been photo-ionised and heated by UV photons \citep{Bolton_2012, Gaikwad_2020}.   These data are consistent with $f_{\rm X}\simeq 10$. Adopting larger X-ray efficiencies in our model would overheat the low density IGM by $z=6$.


\section{The 21 cm forest optical depth}\label{sec:21cm_modeling}

We now turn to the calculation of the \21cm optical depth.  The \21cm line arises from the hyperfine structure of the hydrogen atom, and is determined by the relative orientation of the proton and electron spin, where the ground state energy level is split into a singlet and triplet state.  A photon with rest-frame wavelength $\lambda_{21}=21.11\rm\,cm$, or equivalently frequency $\nu_{21}=1420.41\rm\,MHz$, can induce a transition between these two states.   

In the absence of redshift space distortions, the optical depth to $21\rm\,cm$ photons at redshift $z$ is
\begin{align} \tau_{21}(z) =~& \frac{3h_{\rm p}c^{3}A_{10} }{32\pi \nu_{21}^{2}k_{\rm B}}\frac{n_{\rm HI}(z)}{T_{\rm S}(z)H(z)}, \label{eq:tau21_novpec} \\
  =~ & 0.27 x_{\rm HI}\left(\frac{1+\delta}{10}\right)\left(\frac{T_{\rm S}}{10\rm \,K}\right)^{-1}\left(\frac{1+z}{10}\right)^{3/2}, \nonumber  \end{align}
\noindent
where $n_{\rm HI}$ is the \HI\ number density, $T_{\rm S}$ is the spin temperature, $A_{10}=2.85\times 10^{-15}\rm\,s^{-1}$ is the Einstein spontaneous emission coefficient for the hyperfine transition, $\delta$ is the gas overdensity and $H(z)$ is the Hubble parameter \citep{Madau_1997}. Note the factor of $0.27$ in the second equality is cosmology dependent. Absorption will therefore be most readily observable for dense, cold and significantly neutral hydrogen gas.  The \HI\ spin temperature, a measure of the relative occupation numbers of the singlet and triplet states, is \citep{Field_1958}
\begin{equation}\label{eq:TS}
T_{\rm S}^{-1}= \frac{T^{-1}_{\rm CMB} + x_{\alpha}T^{-1}_{\alpha}+x_{\rm c}T^{-1}_{\rm K}}{1+x_{\alpha}+x_{\rm c}}, \end{equation}
\noindent
where $T_{\rm CMB}=2.73(1+z)\rm\,K$ is the temperature of the cosmic microwave background \citep[CMB,][]{Fixsen_2009}, $T_{\alpha}$ is the \Lya colour temperature and $x_{\rm c}$, $x_{\alpha}$ are the coupling coefficients for collisions and \Lya photon scattering, respectively. If $x_{\rm c}+x_{\alpha}\gg 1$, the \HI\ spin temperature is coupled to the gas kinetic temperature, and if $x_{\rm c}+x_{\alpha}\ll 1$ it is coupled to the CMB temperature.

The collisional coupling coefficient is
\begin{equation} x_{\rm c}=\frac{T_{\star}}{A_{10}T_{\rm CMB}}(\kappa_{10}^{\rm HH}n_{\rm H} + \kappa_{10}^{\rm eH}n_{\rm e} + \kappa_{10}^{\rm pH}n_{\rm p}), \end{equation}
\noindent
where $T_{\star}=h_{\rm p}\nu_{21}/k_{\rm B}$, and $\kappa_{10}^{\rm HH}$, $\kappa_{10}^{\rm eH}$, $\kappa_{10}^{\rm pH}$ are the temperature dependent de-excitation rates for collisions between hydrogen atoms, electrons and hydrogen atoms, and protons and hydrogen atoms, respectively.  We use the convenient fitting functions to the de-excitation rates from \citet{Kuhlen_2006} and \citet{Liszt_2001}, modified to better agree with tabulated values for $\kappa_{10}^{\rm HH}$ \citep{Furlanetto_Oh_2006}, $\kappa_{10}^{\rm eH}$ \citep{Furlanetto_2007_electron}, and $\kappa_{10}^{\rm pH}$ \citep{Furlanetto_2007_proton} over the range $1\rm\,K\leq T_{\rm K} \leq 10^{4}\rm\,K$.

The coupling coefficient for \Lya scattering is \citep{Wouthuysen_1952,Field_1958,Madau_1997} 
\begin{equation} x_{\rm \alpha} = \frac{2T_{\star}\lambda^{3}_{\alpha}\Lambda_{\alpha}}{9T_{\rm CMB}A_{10}h_{\rm p}c}S_{\alpha}J_{\alpha}, \end{equation}
\noindent
where $\lambda_{\alpha}=1215.67$ \AA, $\Lambda_{\alpha}=6.265\times 10^{8}\rm\, s^{-1}$ is the Einstein spontaneous emission coefficient for the \Lya transition, $S_{\alpha}$ is a factor of order unity that corrects for the spectral distortions in the \Lya spectrum, and $J_{\alpha}$ is the proper \Lya specific intensity in units $\rm erg\,s^{-1}\,cm^{-2}\,Hz^{-1}\,sr^{-1}$. We use the fits provided by \citet{Hirata_2006} to calculate $T_{\alpha}$ and $S_{\alpha}$, where $T_{\rm S}$, $T_{\alpha}$ and $S_{\alpha}$ must be solved for iteratively.

In this work we also include the effect of redshift space distortions on the \21cm forest absorption features.  In our calculation of the \21cm optical depth, we therefore include a convolution with the Gaussian line profile and incorporate the gas peculiar velocities from our hybrid RT/hydrodynamical simulations. The optical depth in Eq.~(\ref{eq:tau21_novpec}) may then be calculated in discrete form as \citep[e.g.][]{Furlanetto_2002}
\begin{align}
\tau_{\rm 21, i} =~& \frac{3h_{\rm p}c^{3}A_{10} }{32\pi^{3/2}\nu_{21}^{2}k_{\rm B}} \frac{\delta v}{H(z)} \nonumber \\
             & \times \sum_{j=1}^{N}\frac{n_{{\rm HI}, j}}{b_{j}T_{{\rm S}, j}}\exp\left( - \frac{ (v_{{\rm H},i}-u_{j})^{2}} {b^{2}_{j}}\right), \label{eq:tau21z}
\end{align}
\noindent
for pixel $i$ with Hubble velocity $v_{{\rm H},i}$ and velocity width\footnote{Note the width of the pixel must be smaller than the typical thermal width of an absorber, $\Delta\nu_{21}=0.61(T_{\rm K}/10^{2}\rm\,K)^{1/2}\rm\,kHz$, to ensure the optical depths obtained using Eq.~(\ref{eq:tau21z}) are converged.  In this work we resample the simulation outputs using linear interpolation to achieve the required pixel size.  Alternatively, the line profile may be evaluated using error functions \citep{Meiksin_2011,Hennawi_2020}.}  $\delta v$. Here $b=(2k_{\rm B}T_{\rm K}/m_{\rm H})^{1/2}$ is the Doppler parameter, $T_{\rm K}$ is the gas kinetic temperature, $u_{j}=v_{{\rm H},j}+v_{{\rm pec},j}$, and $v_{\rm pec}$ is the peculiar velocity of the gas.  We evaluate Eq.~(\ref{eq:tau21z}) in our simulations by extracting a total of $5000$ periodic lines of sight, drawn parallel to the simulation box axes at redshift intervals of $\Delta z = 0.1$ over the range $5\leq z\leq 12$.  The total path length we use to make our mock \21cm forest spectra at each output redshift is therefore $200h^{-1}\rm\,cGpc$. 

The redshift evolution of the transmission, $F=e^{-\tau_{21}}$, for a random selection of \21cm forest spectra drawn from the zr53 simulation is shown in Fig.~\ref{fig:long_spectrum}, for three different X-ray efficiencies.  No instrumental features have been added to the simulated data.  The detailed small-scale structure of the \21cm absorption is displayed in the insets. One can see the strong effect that X-ray heating has on the the \21cm\ absorption as the X-ray efficiency parameter is increased from $f_{\rm X}=0.01$ in the top panel, to $f_{\rm X}=1$ in the bottom panel \citep[cf.][]{Xu_2011,Mack_2012}. The redshift evolution due to the increasing filling factor of warm ($T_{\rm K}\sim 10^{4}\rm\,K$), photo-ionised gas is also apparent. In particular, the occurrence of gaps in the \21cm forest absorption due to extended regions of ionised gas increases toward lower redshift. 

In order to better identify the gas associated with the absorption, we calculate the optical depth weighted density, $\Delta_{\rm w}=1+\delta_{\rm w}$, and optical depth weighted kinetic temperature, $T_{\rm K,w}$, for each pixel in our zr53 mock spectra for $f_{\rm X}=0.1$.  This is analogous to the approach used to study the properties of gas responsible for absorption in the \Lya forest \citep{Schaye_1999}; peculiar motions (and to a much lesser extent, line broadening) would otherwise distort the mapping between \21cm optical depth, temperature and gas density. The results are shown in Fig.~\ref{fig:strong_absorbers}, where the temperature-density plane is displayed for the zr53 simulation at three different redshifts: $z=9$ (top), 7.5 (middle) and 6 (bottom).   The colour bar and contours show the average \21cm optical depth and the relative number density of the pixels, respectively.

The gas distribution in Fig.~\ref{fig:strong_absorbers} is bimodal, with the bulk of the pixels associated with either warm ($T_{\rm K}\sim 10^{4}\rm \,K$), photo-ionized gas or cold ($T_{\rm K}\leq 10^{2}\rm \,K$), significantly neutral regions \citep[see also][]{Ciardi_2013,Semelin_2016}.  The plume of gas at intermediate temperatures has been heated by shocks from structure formation.  Note, furthermore, that in this very late reionisation model the IGM is still not fully ionised by $z=6$.  The largest optical depths in the model arise not from the highest density gas, but the cold, diffuse IGM with $3<\Delta<10$. This is because gas at higher densities is typically reionised early due to proximity to the ionising sources, and also because gas around haloes (with $\Delta \gtrsim 100$) is shock-heated and partially collisionally ionised.  Note again, however, there is no cold, star forming gas in this simulation -- for further discussion of this point see Appendix~\ref{sec:starformation}.    Toward lower redshift, the increase in the minimum kinetic temperature of the neutral gas due to X-ray heating, the partial ionisation of the \HI\ by secondary electrons and collisions, and the decrease in the proper number density of gas at fixed overdensity, all conspire to lower the maximum optical depth.  The contours furthermore show that the regions with the largest optical depths are at least 100 times rarer than the bulk of the cold, neutral gas.  Nevertheless, in this very late reionisation model, it remains possible that some detectable \21cm absorption may persist as late as $z\simeq 6$.  We now explore this possibility in more detail.


\section{The detectability of 21 cm forest absorption for very late reionisation}\label{sec:results}

\begin{figure}
    \begin{minipage}{\columnwidth}
 	  \centering
 	  \includegraphics[width=\linewidth]{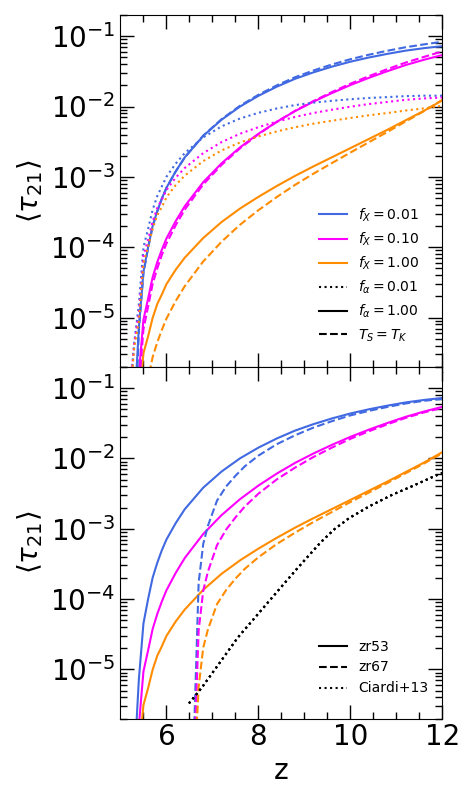}
	\end{minipage}
	\vspace{-0.5cm}
    \caption{Redshift evolution of the volume averaged 21 cm optical depth in the zr53 model (solid curves) for a \Lya efficiency $f_{\alpha}=1$ and an assumed X-ray efficiency of $f_{\rm X}=0.01$ (blue curves), $0.1$ (fuchsia curves) and $1$ (orange curves).  In the upper panel, this is compared to results from the same hybrid RT/hydrodynamical simulation, but with 21 cm optical depths calculated under the assumption of strong (i.e. $T_{\rm S}=T_{\rm K}$, shown by the dashed curves) and weak \Lya coupling ($f_{\alpha}=0.01$, shown by the dotted curves).   In the lower panel, the dashed curves instead show $\langle \tau_{21} \rangle$ for the hybrid model with an earlier end to reionisation (zr67). The dotted black curve in the lower panel corresponds to the RT+Ly$\alpha$+x model from fig. 2 of \citet{Ciardi_2013}.}
    \label{fig:tau_evo}
\end{figure}

\begin{figure*}
    \begin{minipage}{2\columnwidth}
 	  \centering
 	  \includegraphics[width=\linewidth]{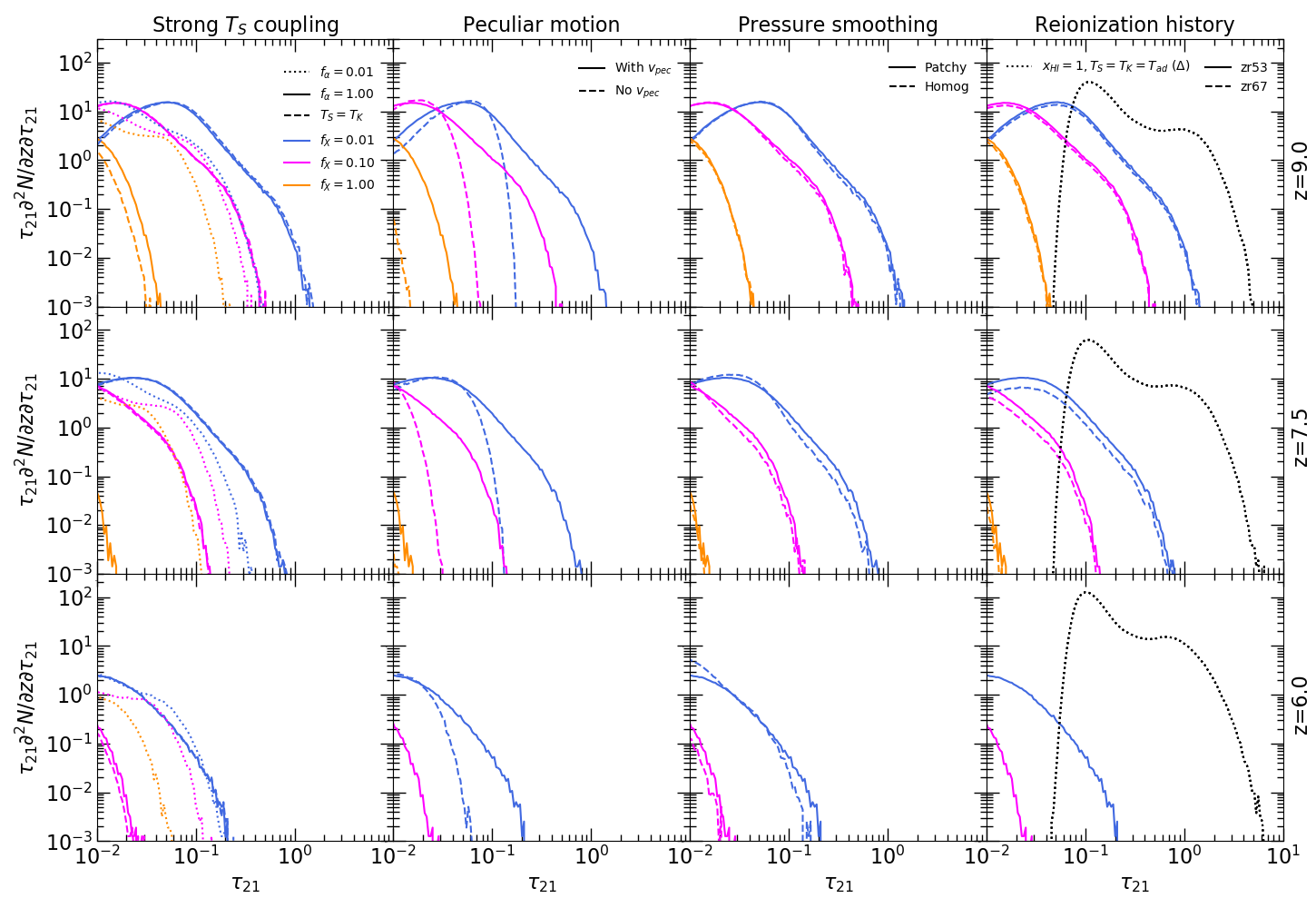}
	\end{minipage}
	\vspace{-0.3cm}
    \caption{The differential number density of absorption lines in synthetic \21cm forest spectra. Each row shows the distribution at redshift $z=9$ (top row), $7.5$ (middle row) and $6$ (bottom row) for our fiducial model with \Lya efficiency $f_{\alpha}=1$, and in each panel the distribution is shown for three X-ray efficiencies, $f_{\rm X}=0.01$ (blue curves),$f_{\rm X}=0.1$ (fuchsia curves) and $f_{\rm X}=1$ (orange curves). Each column displays the zr53 simulation (solid curves) compared to models where one of the parameter choices is varied (dashed curves).  These parameters are, from left to right, the assumption of strong \Lya coupling (i.e. $T_{\rm S}=T_{\rm K}$), neglecting the effect of peculiar velocities (i.e $v_{\rm pec}=0$), pressure smoothing due to a uniform rather than patchy UV photo-heating rate (i.e the zr53-homog model) and an earlier end to reionisation (the zr67 model).  In the first column, we also show the number density distribution for very weak \Lya coupling (i.e. $f_{\alpha}=0.01$, dotted curves).  The black dotted curves in the last column show the case of no reionisation or X-ray heating (i.e. $T_{\rm S}=T_{\rm K}=T_{\rm ad}=2.73\rm\, K\ (1+\delta)^{2/3}(1+z)^2/(1+z_{\rm dec})$, where $z_{\rm dec}=147.8$ \citep{Furlanetto_Oh_2006}, and $x_{\rm HI}=1$).} 
    \label{fig:AoFdiff}
\end{figure*}

\subsection{The volume averaged 21 cm optical depth}\label{sec:tau_ave}

We first consider the redshift evolution of the volume averaged \21cm optical depth, $\langle \tau_{21} \rangle$, in the zr53 simulation, displayed as the solid curves in Fig.~\ref{fig:tau_evo} for our fiducial model with $f_{\alpha}=1$.  In the upper panel, we test the common assumption that, as a result of the Wouthuysen-Field effect, the spin temperature becomes strongly coupled to the gas kinetic temperature during the later stages of reionisation, such that $T_{\rm S}=T_{\rm K}$ \citep[e.g.][]{Xu_2009,Mack_2012,Ciardi_2013}.  This is shown by the dashed curves in the upper panel of Fig.~\ref{fig:tau_evo}.  As also noted by \citet{Semelin_2016}, a full calculation of $T_{\rm S}$ using Eq.~(\ref{eq:TS}) can either reduce or enhance \21cm optical depths relative to the value obtained assuming strong coupling.  This is caused by a partial coupling of the spin temperature to the CMB temperature; if $T_{\rm K}<T_{\rm CMB}$, the full calculation will result in a higher spin temperature and smaller \21cm optical depth, and vice versa. 

This can be observed in Fig.~\ref{fig:tau_evo} for $f_{\rm X}=0.1$ (fuchsia curves), where $\langle \tau_{21}\rangle$ for the full calculation assuming $f_{\alpha}=1$ (solid curves) is smaller than the $T_{\rm S}=T_{\rm K}$ case (dashed curves) at $z\gtrsim 8$, but is greater at lower redshifts.  This coincides with the temperature evolution shown in Fig.~\ref{fig:Tevol}, particularly the transition from $T_{\rm K}<T_{\rm CMB}$ (and $T_{\rm S}>T_{\rm K}$) at $z>8$ to $T_{\rm K}>T_{\rm CMB}$ (and $T_S<T_K$) at $z<8$.  Similarly, in the case of a weaker ($f_{\rm X}=0.01$, blue curves) or stronger ($f_{\rm X}=1$, orange curves) X-ray background, the full $T_{\rm S}$ calculation respectively decreases or increases $\langle \tau_{21}\rangle$ relative the the strong coupling approximation.   The dotted curves furthermore show the $\langle \tau_{21} \rangle$ redshift evolution for significantly weaker \Lya coupling, with $f_{\alpha}=0.01$. In this case $T_{\rm S}$ is now decoupled from $T_{\rm K}$ and has a value similar to $T_{\rm CMB}$.   The weak coupling means $\langle \tau_{21} \rangle$ is significantly increased in the models with efficient X-ray heating.  Hence, while the assumption of strong coupling, $T_{\rm S}=T_{\rm K}$, remains a reasonable approximation if $f_{\alpha}=1$, this will not be the case if the background \Lya emissivity is significantly overestimated in our fiducial model (i.e. $f_{\alpha} \ll 1$).

The lower panel of Fig.~\ref{fig:tau_evo} instead shows $\langle \tau_{21} \rangle$ for the two different reionisation histories in Fig.~\ref{fig:intro_sim}. Both of these reionisation models are broadly consistent with existing constraints on the timing of reionisation, and the zr53 model furthermore successfully reproduces the large fluctuations in the \Lya forest opacity at $z=5.5$ \citep{Kulkarni_2019}.   For comparison, we also show $\langle \tau_{21} \rangle$ from \citet{Ciardi_2013} as the dotted curve.  This includes X-ray and \Lya heating following \citet{Ciardi_2010}, and is most similar to our zr67 simulation with $f_{\rm X}\simeq 1$.  The differences between this work and \citet{Ciardi_2013} are due to different assumptions for the X-ray emissivity and the reionisation history.  A later end to reionisation means $\langle \tau_{21} \rangle$ in Fig.~\ref{fig:tau_evo} remains significantly larger than earlier reionisation models at redshifts $6\lesssim z \lesssim 7$.  If reionisation does indeed complete late, such that large neutral islands persist in the IGM at $z\simeq 6$ \citep[e.g.][]{Lidz_2007,Mesinger_2010}, this suggests \21cm forest absorption lines may be more readily observable than previously thought at these redshifts. 

\subsection{The differential number density of 21 cm absorption lines} \label{sec:diffdist}

We now consider the number density of individual absorption lines in our high resolution mock spectra.  We present this as the total number of lines, $N$, within a given optical depth bin, per unit redshift \citep[see also][]{Furlanetto_2006a,Shimabukuro_2014}, where
\begin{equation} f(\tau_{21},z)=\frac{\partial^{2}N}{\partial \tau_{21}\partial z}. \end{equation}
The absorption lines in our simulated \21cm forest spectra are identified following a similar method to \citet{Garzilli_2015}, who identify absorption lines in mock \Lya forest spectra as local optical depth maxima located between two minima.  In this work, we require that the local maxima must have a prominence (i.e. be higher by a certain value than the minima) that corresponds to a factor of 1.001 difference in the transmitted flux, $F=e^{-\tau_{21}}$, between the line base and peak.  We then define the optical depth for each identified line as being equal to the local maximum.  We find this method is robust for lines with $\tau_{21}\geq 10^{-2}$ (i.e. $F=e^{-\tau_{21}}\leq 0.99$), but for optical depths below this threshold the number of lines is sensitive to the choice for the prominence, and is thus unreliable.

The number density distributions, $\tau_{21}f(\tau_{21},z)$, for different model parameters at three different redshifts, $z=9$, $7.5$ and $6$, are displayed for our fiducial model with $f_{\alpha}=1$ in Fig.~\ref{fig:AoFdiff} (for an illustration of the effect of these model parameter variations on individual absorbers, see Appendix~\ref{sec:specparam}).  Each column corresponds to a different model parameter choice, each row shows a different redshift, and in each panel we show the distribution for three X-ray efficiencies: $f_{\rm X}=0.01$ (blue curves), $f_{\rm X}=0.1$ (fuchsia curves) and $f_{\rm X}=1$ (orange curves). The peak of the distribution is at $\tau_{21}\leq 0.1$, and it shifts to lower amplitudes and smaller optical depths as the IGM reionises and the spin temperature of the X-ray heated gas increases.  The distribution also has an extended tail toward higher optical depths.  While strong \21cm absorbers will be rare, this suggests that for $f_{\rm X}\sim 0.1$, features with a transmission of $F=e^{-\tau_{21}}\simeq 0.9$ should still be present at $z=7.5$ in the late reionisation model (see also Fig.~\ref{fig:long_spectrum}). 

In the first column of Fig.~\ref{fig:AoFdiff} we re-examine the effect of strong \Lya coupling on the distribution of \21cm optical depths.  As was the case for the volume averaged optical depth in Fig.~\ref{fig:tau_evo}, the impact is relatively modest for low X-ray efficiencies: for $f_{\rm X}=0.01$  at $z=6$, the two cases are almost identical.   For $f_{\rm X}=1$, however, the abundance of features with $\tau_{21}\geq0.01$ for $T_{\rm S}=T_{\rm K}$ is more than 50 per cent smaller than the full calculation at $z=9$.  In either case, however, by $z=7.5$ most gas in the $f_{\rm X}=1$ model has $\tau_{21}<10^{-2}$, and will therefore be challenging to detect directly.  However, the dotted curves also demonstrate that if $f_{\alpha}=0.01$, the weak coupling of $T_{\rm S}$ to $T_{\rm K}$ allows strong \21cm absorbers to still be observable at $z=6$, even for $f_{\rm X}=1$. 

We consider the effect of gas peculiar velocities on the \21cm forest in the second column of Fig.~\ref{fig:AoFdiff}.  Redshift space distortions are well known to impact on the observability of the high redshift \21cm signal \citep{Bharadwaj_2004,Mao_2012,Majumdar_2020}.  We do this by creating mock \21cm spectra that ignore the effect of gas peculiar motions, such that $v_{\rm pec}=0$ in Eq.~(\ref{eq:tau21z}).  The results are shown by the dashed curves.  While the position of the peak in the number density distribution is unchanged, the high optical depth tail is strongly affected, particularly for inefficient X-ray heating.  Ignoring peculiar velocities within 21 cm forest models can therefore significantly reduce the incidence of the strongest absorbers, and this will have a negative impact on the predicted observability of the \21cm forest.  Qualitatively, this agrees with the assessment of \citet{Semelin_2016}, who also included the effect of gas peculiar motions in their models.

As our hybrid simulations self-consistently model the hydrodynamical response of gas to photo-heating by the inhomogeneous UV radiation field, we may also estimate the effect of (the lack of) pressure (Jeans) smoothing on the \21cm forest.  Inhomogeneous reionisation introduces large scale gas temperature fluctuations in the IGM \citep{Keating_2018}, and these lead to differences in the local gas pressure that smooth the structure of the IGM on different scales \citep[e.g.][]{Gnedin_1998,Kulkarni_2015,Nasir_2016,DAloisio_2020}.  In the absence of significant X-ray heating, the neutral gas responsible for the \21cm forest should therefore experience minimal pressure smoothing compared to the photo-ionised IGM.  We therefore compare the results of our zr53 model to the zr53-homog simulation in the third column of Fig.~\ref{fig:AoFdiff}.  The latter model has exactly the same initial conditions and volume averaged reionisation history as zr53, but all the gas in the simulation volume is instead heated simultaneously (i.e. we do not follow the radiative transfer for UV photons).  

The dashed curves in the third column of Fig.~\ref{fig:AoFdiff} show the line density distribution obtained from the density and peculiar velocity fields in the zr53-homog model (differences due to $x_{\rm HI}$, $T_{\rm K}$ and $T_{\rm S}$ in the two models have been removed).  We observe that there is a small, redshift dependent difference between the two distributions, such that the simulation with the homogeneous UV background exhibits fewer strong absorption lines.  This is because the gas responsible for the highest optical depths in the \21cm forest (see Fig.~\ref{fig:strong_absorbers}) is still cold within the hybrid model, and hence has slightly higher density due to the smaller pressure smoothing scale.  

We caution, however, that this comparison will still not fully capture the effect of pressure smoothing on \21cm forest absorbers.  For reference, the comoving pressure smoothing scale in the IGM is \citep{Gnedin_1998,Garzilli_2015}
\begin{align} \lambda_{\rm p} =&~ f_{\rm J}\frac{\lambda_{\rm J}}{2\pi} = f_{\rm J} \left(\frac{10 k_{\rm B}T_{\rm K}}{9\mu m_{\rm H} (1+\delta)  H_{0}^{2}\Omega_{\rm m}(1+z)}\right)^{1/2}, \nonumber \\
  =&~ 1.5h^{-1}\rm\,ckpc \, f_{\rm J}  \left[\left(\frac{10}{1+\delta}\right)\left(\frac{1.22}{\mu}\right)\left(\frac{T_{\rm K}}{10^{2}\rm\,K}\right)\left(\frac{1+z}{10}\right)\right]^{1/2}, \label{eq:Jeans} \end{align}
where $\lambda_{\rm J}$ is the Jeans scale, $\mu$ is the mean molecular weight of hydrogen and helium assuming primordial composition ($\mu=1.22$ for fully neutral gas, $\mu=0.59 $ for fully ionised), and $f_{\rm J}=\lambda_{\rm p}/\lambda_{\rm J}$ is a factor of order unity that accounts for the finite time required for gas to dynamically respond to a change in pressure.  For comparison, the mean interparticle separation and gravitational softening length in our simulations are $19.5h^{-1}\rm\,ckpc$ and $0.78h^{-1}\rm\,ckpc$, respectively.  Eq.~(\ref{eq:Jeans}) thus implies that the pressure smoothing scale for typical \21cm forest absorbers is not fully resolved in our simulations \citep[see also][]{Emberson_2013}.  We furthermore do not capture the \21cm absorption from minihaloes with $M<2.5\times 10^{7}\,M_{\odot}$ \citep{Furlanetto_2006a}.  Larger differences could then be observed in Fig.~\ref{fig:AoFdiff} for fully resolved gas.  On the other hand, although we follow the dynamical response of gas to heating by UV photons, the X-ray heating of the neutral gas in our hybrid simulation is applied in post-processing.  It is therefore decoupled from the hydrodynamics, and this may then underestimate the impact of pressure smoothing on cold gas for high X-ray efficiencies.   Regardless of these modelling uncertainties, however, this suggests that the effect of the pressure smoothing scale on the \21cm forest in the diffuse IGM remains small compared to the substantial impact of X-ray heating on the spin temperature at $z\leq 10$.

Finally, in the fourth column of Fig.~\ref{fig:AoFdiff} the effect of the reionisation history is displayed for the zr53 (solid curves) and zr67 (dashed curves) simulations.  For comparison, the dotted curves also show the line number density distribution under the assumption of no reionisation or X-ray heating (i.e. $T_{\rm S}=T_{\rm K}=T_{\rm ad}$ and $x_{\rm HI}=1$).    As expected, the two reionisation models are significantly different at $z=6$;  there are no strong absorption features with $\tau_{21}>10^{-2}$ in zr67 model, as reionisation has already completed by this time.  At $z=7.5$ one can see that there are also fewer absorption features in the zr67 model due to the larger volume of ionised gas.  However, the differences between the two models become smaller with increasing redshift.   This again demonstrates that for reionisation models that complete at $z<6$, the \21cm forest may remain observable if sufficiently bright radio sources exist at $6<z<7$.   Alternatively, a null-detection could place an interesting limit on the very uncertain X-ray background \citep[e.g.][]{Mack_2012}.  We now investigate this possibility further.

\subsection{Detectability of strong 21 cm forest absorbers at redshift \texorpdfstring{$z=6$}{} for late reionisation and X-ray heating}\label{sec:observability}

\begin{figure*}
    \begin{minipage}{2\columnwidth}
 	  \centering
 	  \includegraphics[width=\linewidth]{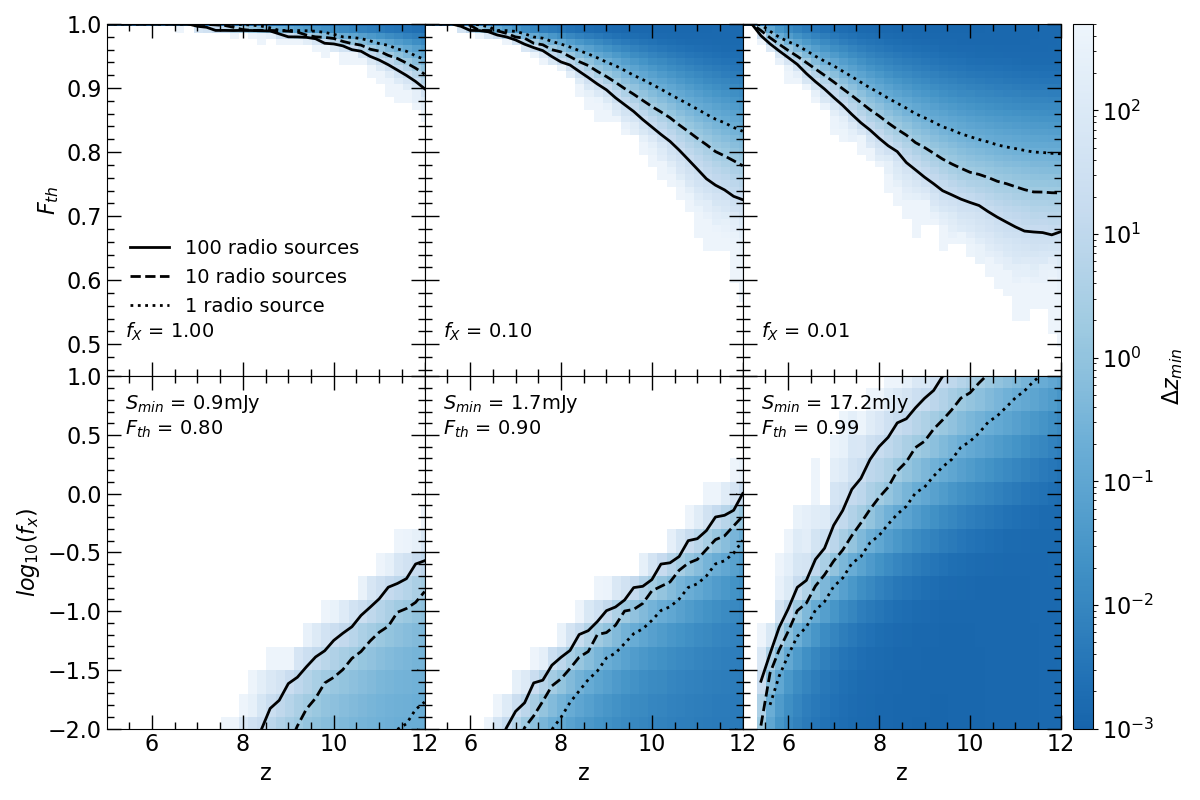}
	\end{minipage}
	\vspace{-0.3cm}
    \caption{The minimum redshift path length, $\Delta z_{\rm min}$, required to observe a single $21\rm\,cm$ absorption feature in the zr53 simulation \emph{assuming that a sufficient number of background radio sources exist.}    The mock spectra have been convolved with a boxcar of width $5 \rm\, kHz$ to approximately model the effect of spectral resolution on the lines.  In the upper panels we show $\Delta z_{\rm min}$ in the $F_{\rm th}$--z plane for an X-ray efficiency factor of $f_{\rm X}=1$ (left), $0.1$ (middle) and $0.01$ (right). In the lower panels we instead show $\Delta z_{\rm min}$ in the $f_{\rm X}$--z plane for a \21cm  absorption feature with minimum transmission  $F\leq F_{\rm th}=0.8$ (left), $0.9$ (middle) and $0.99$ (right). Here we also note the minimum intrinsic flux density, $S_{\rm min}$, that a background radio source must have such that an absorption line with a minimum at $F\leq F_{\rm th}$ is detectable with SKA1-low at a signal-to-noise of $\rm S/N=5$ and integration time of $t_{\rm int}=1000\rm\,hr$ (see Eq.~\ref{eq:radiometer}). The unshaded white regions are where no absorbers are present over our total simulated path length of $200h^{-1}\rm\,cGpc$.  The thick black curves in each panel track the redshift path length that would be covered by the observation of 1 (dotted), 10 (dashed) and 100 (solid) radio sources assuming redshift bins of width $\Delta z=0.2$. }
    \label{fig:obs_fX}
\end{figure*}

\begin{figure*}
    \begin{minipage}{2\columnwidth}
 	  \centering
 	  \includegraphics[width=\linewidth]{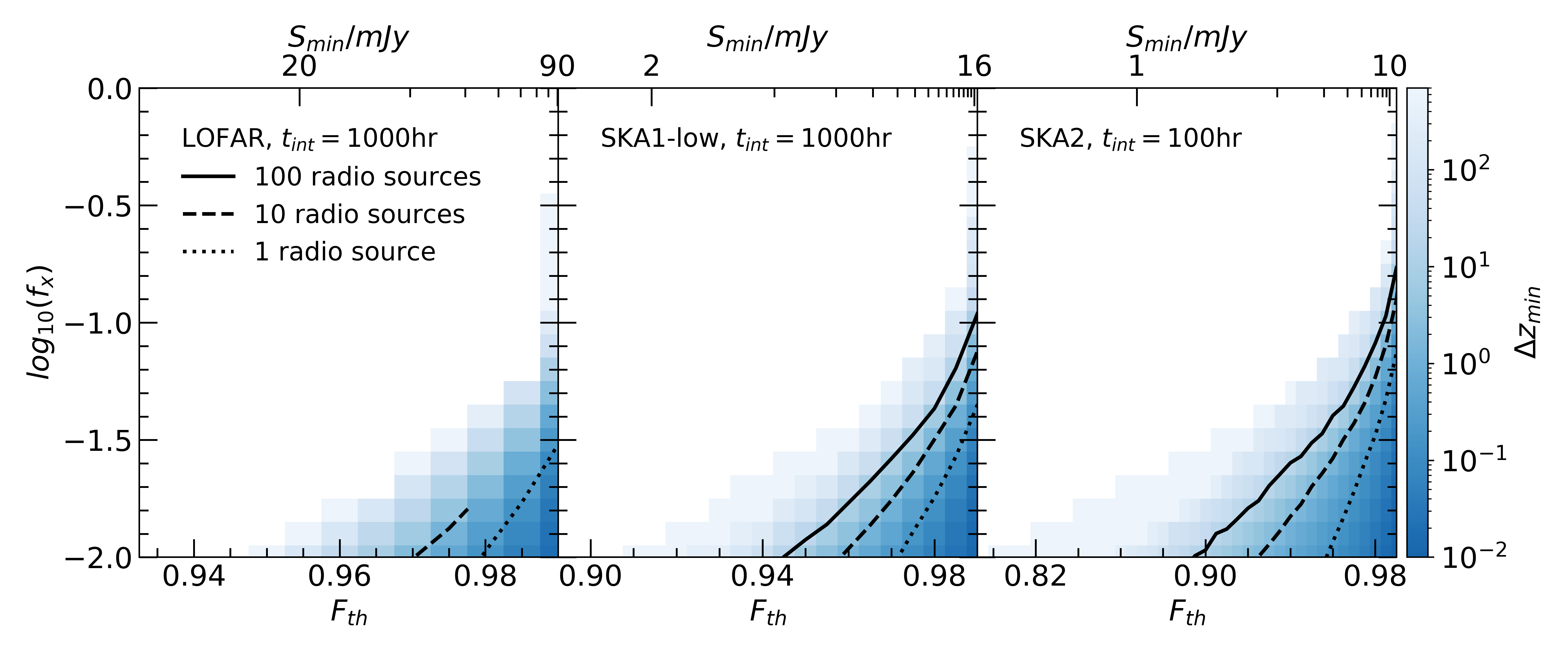}
	\end{minipage}
	\vspace{-0.3cm}
    \caption{As for Fig.~\ref{fig:obs_fX}, but now the minimum redshift path length $\Delta z_{\rm min}$ is shown in the $f_{\rm X}$--$F_{\rm th}$ plane at redshift $z=6$.    The mock spectra have been convolved with a boxcar of width $10 \rm\, kHz$ (left), $5 \rm\, kHz$ (middle) and $1 \rm\, kHz$ (right) to approximately model the effect of our assumed bandwidths for LOFAR, SKA1-low and SKA2, respectively.  Note the scale on the horizontal axis is different in each panel. The upper horizontal axis now also shows the minimum intrinsic flux density, $S_{\rm min}$, required for a background radio source, such that a line with minimum transmission $F\leq F_{\rm th}$ is detectable at a signal-to-noise of $\rm S/N=5$ by LOFAR with an integration time of $t_{\rm int}=1000\rm\,hr$ (left), by SKA1-low with $t_{\rm int}=1000\rm\,hr$ (middle) and by SKA2 with $t_{\rm int}=100\rm\,hr$ (right).  In the left panel (LOFAR) the thick dashed curve is truncated where the number of available background radio sources predicted by \citet{Saxena_2017} with $S_{\rm min}$ at $z\simeq 6$ falls below $10$.  The curve for $100$ sources (solid) is not shown, as this exceeds the expected radio source number from \citet{Saxena_2017} at the required $S_{\rm min}$.} 
    \label{fig:obs_fXvsFth}
\end{figure*}

A detection of the \21cm forest relies on the identification of objects at high redshift that are sufficiently radio bright to act as background sources.  Based on a model for the radio galaxy luminosity function at $z>6$, \citet{Saxena_2017} predict around one radio source per 400 square degrees at a flux density limit of $S_{150\rm\,MHz}=3.5\rm\,mJy$, and at least $\sim 30 $ bright sources with  $S_{150\rm\,MHz}>15\rm\,mJy$ \citep[see also][]{Bolgar_2018}.  Ongoing observational programmes such as the LOFAR Two-metre Sky Survey  \citep[LoTSS,][]{Shimwell_2017,Kondapally_2020}, the Giant Metrewave Radio Telescope (GMRT) all sky radio survey at $150\rm\,MHz$ \citep{Intema_2017}, and the Galactic and Extragalactic All-sky Murchison Widefield Array survey \citep[GLEAM,][]{Wayth_2015} should furthermore detect hundreds of bright $z>6$ radio sources.  Encouragingly, a small number of radio-loud sources have already been identified at $z>5.5$ 
\citep[e.g.][]{Banados_2018}, including the $z=6.1$ blazar PSO J0309+27 with a flux density $S_{147\rm\,MHz}=64.2\pm 6.2\rm\,mJy$ \citep{Belladitta_2020}

We now use our hydrodynamical simulations to assess the feasibility of detecting the \21cm forest in late reionisation models, assuming
$f_{\alpha}=1$.  We shall calculate the minimum redshift path length, $\Delta z_{\rm min}$, necessary for detecting a single, strong (i.e. $\tau_{21}>0.01$) absorption line with a minimum transmission at some arbitrary threshold $ F_{\rm th}=e^{-\tau_{21,\rm\,th}}$.  For a signal-to-noise ratio $\rm S/N$, the minimum flux density contrast, $\Delta S_{\rm min}$, detectable by an interferometric radio array is then \cite[e.g.][]{Ciardi2015_SKA}, 
\begin{equation} \Delta S_{\rm min} = S_{\rm min}-S_{\rm abs}= \frac{2k_{\rm B}T_{\rm sys}}{A_{\rm eff} \sqrt{\Delta \nu t_{\rm int}}} \rm\, S/N , \label{eq:radiometer} \end{equation}
where $T_{\rm sys}$ is the system temperature, $\Delta \nu$ is the bandwidth, $A_{\rm eff}$ is the effective area of the telescope, $t_{\rm int}$ is the integration time, and $S_{\rm min}$ is the minimum intrinsic flux density a radio source must have to allow detection of a \21cm absorption feature with a minimum at a flux density of $S_{\rm abs}=S_{\rm min}e^{-\tau_{21,\rm\,th}}$.  Adopting some representative values in Eq.~(\ref{eq:radiometer}), the minimum flux density required to detect a \21cm absorption feature with a minimum transmission $F_{\rm th}$ is therefore
\begin{align} S_{\rm min} =~& 10.3\rm\,mJy \left(\frac{0.01}{1-F_{\rm th}}\right)\left(\frac{\rm S/N}{5}\right)\left(\frac{5\rm\,kHz}{\Delta \nu}\right)^{1/2}\left(\frac{1000\rm\,hr}{t_{\rm int}}\right)^{1/2} \nonumber \\ &\times \left(\frac{1000\rm\,m^{2}\,K^{-1}}{A_{\rm eff}/T_{\rm sys}}\right). \end{align} 
\noindent
In what follows, we shall adopt values for the sensitivity, $A_{\rm eff}/T_{\rm sys}$, in Eq.~(\ref{eq:radiometer}) appropriate for LOFAR, SKA1-low and SKA2, where $A_{\rm eff}/T_{\rm sys}\simeq 80\rm\, m^{2}\,K^{-1}$, $600\rm\, m^{2}\,K^{-1}$ and $5500\rm\, m^{2}\,K^{-1}$, respectively\footnote{Note that in reality the sensitivity $A_{\rm eff}/T_{\rm sys}$ is frequency dependent.  However, over the frequency range we consider, $142\rm\,MHz\leq \nu_{21}/(1+z) \leq 203\rm\,MHz$, this dependence is reasonably weak.  See fig. 8 in \citet{Braun_2019} for further details.} \citep{Braun_2019}.   Additionally, to approximately model the effect of spectral resolution on the data we convolve our mock spectra with a boxcar function.  Following the bandwidths adopted in \citet{Ciardi_2015_GRB}, we assume boxcar widths of $10\rm\,kHz$ and $5\rm\,kHz$ for LOFAR and SKA1-low, respectively.  For a more futuristic measurement with SKA2, we assume a smaller bandwidth and adopt a boxcar width of $1\rm\,kHz$.

First, in Fig.~\ref{fig:obs_fX}, we show the minimum redshift path length $\Delta z_{\rm min}$ required to detect a single \21cm absorption line in the minimum transmission threshold $F_{\rm th}$-redshift plane for three different X-ray efficiencies $f_{\rm X}$ (upper panels), or in the $f_{\rm X}$-redshift plane for three different transmission thresholds $F_{\rm th}$ (lower panels).    Note that for now we assume a sufficient number of background radio sources exists for such a measurement; we consider the issue of detectability at $z=6$ further in Fig.~\ref{fig:obs_fXvsFth}.  The mock spectra used in Fig.~\ref{fig:obs_fX} are drawn from the zr53 simulation and have been convolved with a boxcar of width $5\rm\,kHz$ (i.e. our assumed SKA1-low bandwidth).  Unshaded white regions indicate where no absorbers are present over our total simulated path length of $200h^{-1}\rm\,cGpc$.   Fig.~\ref{fig:obs_fX} shows that no absorption features with $F_{\rm th}\leq 0.77$ should be present at $z\lesssim 8$ for even a very low X-ray efficiency of $f_{\rm X}=0.01$ in the late reionisation model. Similarly, almost no strong \21cm absorption with $F_{\rm th} \lesssim 0.99$ will exist at $z<7$ for $f_{\rm X} \geq 1$.  This highlights the challenging nature of \21cm forest measurements from the diffuse IGM, even if reionisation ends very late, and also how sensitive the \21cm forest absorption is to X-ray heating.  Proposals to use the \21cm forest as a sensitive probe for distinguishing between different cosmological or dark matter models using the diffuse IGM are therefore likely to be restricted to very high redshifts, prior to any substantial X-ray heating of the IGM.

As a reference, the black curves in Fig.~\ref{fig:obs_fX} correspond to the redshift path length obtainable by a hypothetical observation of $1$, $10$ or $100$ radio sources of sufficient brightness in redshift bins of width $\Delta z=0.2$  {(i.e. an observation of $N$ radio sources provides a total redshift path length of $0.2N$).}\footnote{The choice of $\Delta z=0.2$ is somewhat arbitrary – we require a bin that is small enough that redshift evolution is not significant, but large enough to probe a reasonable path length.  For reference, increasing the bin size to $\Delta z=0.4$ would approximately halve the number of background sources required to detect a single absorber with $F_{\rm th}$, assuming minimal redshift evolution across the bin.}  A null-detection over this path length would provide a model dependent lower limit on the X-ray background emissivity, such that $f_{\rm X}\geq f_{\rm X,max}$, where $f_{\rm X,max}$ is the maximum X-ray efficiency that retains at least one strong absorption feature with $F\leq F_{\rm th}$. From the lower middle panel in Fig.~\ref{fig:obs_fX}, the null-detection of a feature with $F_{\rm th}<0.9$ at $z=9$ in $1$ ($10$) radio source(s) implies $f_{\rm X,max}\simeq 0.04$ ($f_{\rm X.max} \simeq 0.07$).  The parameter space that lies below the black curves would then be disfavoured.

\begin{table}
	\centering
	\caption{The maximum X-ray background efficiency, $f_{\rm X, max}$, that retains at least one strong \21cm absorption feature with transmission $F\leq F_{\rm th}$ in our synthetic \21cm forest spectra, for a redshift path length corresponding to $N$ bright radio sources covering a redshift bin of width $\Delta z=0.2$, centred at redshift $z=6$.  The mock spectra have been convolved with a boxcar of width $5 \rm kHz$ to approximately model the effect of observed bandwidth on the lines.  The minimum intrinsic flux density of the background source, $S_{\rm min}$, required to detect a line with $F_{\rm th}$ at a signal-to-noise of $\rm S/N=5$ with SKA1-low is calculated using Eq.~(\ref{eq:radiometer}), assuming a bandwidth $\Delta \nu=5\rm\,kHz$, sensitivity $A_{\rm eff}/T_{\rm sys}=600\rm\,m^{2}\,K^{-1}$ and integration time of $t_{\rm int}=1000\rm\,hr$.  The expected number of radio sources in the sky with $S_{\rm min}$ at $z=6$, N$_{\rm S17}$, are estimated from \citet{Saxena_2017} (their fig. 11).  In the event of a null-detection of an absorption feature with $F_{\rm th}$, the $f_{\rm X, max}$ values give a (model dependent) lower limit on the X-ray efficiency.}
	\label{tab:observability}
	\begin{tabular}{c|c|c|ccc}
		\hline
	  & & $z=6$ &\multicolumn{3}{|c|}{$f_{\rm X, max},z=6$} \\
		 	$F_{\rm th}=e^{-\tau_{21\rm,th}}$ & $S_{\rm min}[\rm mJy]$ & N$_{\rm S17}$ & N=1 & N=10 & N=100 \\
		\hline
0.99 & 17.2 & $\sim$100 & 0.045      & 0.075      & 0.109      \\
0.95 & 3.4 & $\sim$2400 & <$10^{-3}$ & 0.007      & 0.012     \\
0.9  & 1.7 & $\sim$6100 & <$10^{-3}$ & <$10^{-3}$ & <$10^{-3}$ \\
	\hline
	\end{tabular}
\end{table}

\begin{table}
	\centering
	\caption{As for Table~\ref{tab:observability}, except the mock $21\rm\,cm$ forest spectra are now smoothed with a boxcar of width $\rm 10\rm\,kHz$ and the minimum source flux densities, $S_{\rm min}$, have been computed for LOFAR using a bandwidth $\Delta \nu=10\rm\,kHz$, sensitivity $A_{\rm eff}/T_{\rm sys}=80\rm\,m^{2}\,K^{-1}$ and integration time of $t_{\rm int}=1000\rm\,hr$. Dashes mean that $f_{\rm X, max}$ is not measurable due to the lack of expected sources.}
	\label{tab:observability2}
	\begin{tabular}{c|c|c|ccc|}
		\hline
	  & & $z=6$ &\multicolumn{3}{|c|}{$f_{\rm X, max},z=6$} \\
		 	$F_{\rm th}=e^{-\tau_{21\rm,th}}$ & $S_{\rm min}[\rm mJy]$ & N$_{\rm S17}$ & N=1 & N=10 & N=100 \\ 
		\hline
0.99 & 91.0 & $\sim 1$         & 0.030      & --- & ---     \\
0.95 & 18.2 & $\sim$90  & <$10^{-3}$ & 0.001      & ---     \\
0.9  & 9.1 & $\sim$420  & <$10^{-3}$ & <$10^{-3}$ & <$10^{-3}$\\
	\hline
	\end{tabular}
\end{table}

\begin{table}
	\centering
	\caption{As for Table~\ref{tab:observability}, except the mock $21\rm\,cm$ forest spectra are now smoothed with a boxcar of width $\rm 1\rm\,kHz$ and the minimum source flux densities, $S_{\rm min}$, have been computed for SKA2 using a bandwidth $\Delta \nu=1\rm\,kHz$, sensitivity $A_{\rm eff}/T_{\rm sys}=5500\rm\,m^{2}\,K^{-1}$ and integration time of $t_{\rm int}=100\rm\,hr$.}
	\label{tab:observability3}
	\begin{tabular}{c|c|c|ccc}
		\hline
	  & & $z=6$ &\multicolumn{3}{|c|}{$f_{\rm X, max},z=6$} \\
		 	$F_{\rm th}=e^{-\tau_{21\rm,th}}$ & $S_{\rm min}[\rm mJy]$ & N$_{\rm S17}$ & N=1 & N=10 & N=100  \\
		\hline
0.99 & 13.2 & $\sim$190 & 0.074      & 0.125      & 0.172   \\
0.95 & 2.6 & $\sim$3600 & 0.007      & 0.020      & 0.031    \\
0.9  & 1.3 & $\sim$8000 & <$10^{-3}$ & 0.004      & 0.011   \\
0.8  & 0.7 & $\sim$5300 & <$10^{-3}$ & <$10^{-3}$ & <$10^{-3}$ \\
	\hline
	\end{tabular}
\end{table}

In practice, however, radio telescope sensitivity, spectral resolution and the availability of sufficiently bright background radio sources will impact upon the detectability of strong lines.  We quantify this in Fig. \ref{fig:obs_fXvsFth}, where similarly to Fig. \ref{fig:obs_fX} we show $\Delta z_{\rm min}$, but now in the $f_{\rm X}$--$F_{\rm th}$ plane at redshift $z=6$.   This is shown for our LOFAR (left), SKA1-low (middle) and SKA2 (right) model assumptions, where we have convolved the synthetic spectra with a boxcar of width $10\rm\, kHz$, $5\rm\, kHz$ and $1\rm\, kHz$, respectively.  The minimum intrinsic source flux density, $S_{\rm min}$, required to detected a line with $F_{\rm th}$ has also been calculated using Eq.~(\ref{eq:radiometer}) and is displayed on the horizontal top axis.  Here we assume a strong absorption line with minimum transmission $F_{\rm th}$ is detected with $\rm S/N=5$ for an integration time of $t_{\rm int}=1000\rm\,hrs$ with LOFAR and SKA1-low, and $t_{\rm int}=100\rm\,hrs$ with SKA2.  First, one can see that if using {a more sensitive telescope with higher spectral resolution it is possible to detect deeper, narrower absorption features}.   Moreover, tighter constraints on the X-ray efficiency $f_{\rm X}$ may also be obtained.  For example, at $z=6$, there are no absorption features with $F\leq0.95$ for $f_{\rm X}>0.01$ if observed by LOFAR.  However, this increases to $f_{\rm X}>0.025$ for SKA1-low and $f_{\rm X}>0.05$ for SKA2.  The minimum source flux density required to detect an absorption feature at fixed $\rm S/N=5$ also decreases significantly, thus increasing the number of potentially suitable background radio sources.


We quantify this in more detail in Tables~\ref{tab:observability}, \ref{tab:observability2}  and \ref{tab:observability3}, where we list the maximum X-ray efficiency, $f_{\rm X, max}$, that retains at least one \21cm absorption feature at $z=6$ with a transmission minimum $F\leq F_{\rm th}$ over a path length of $\Delta z=0.2$, $\Delta z=2$ or $\Delta z=20$ in the zr53 simulation.  This corresponds to $N=1$, $10$ and $100$ sources, respectively, for redshift bins of width $\Delta z=0.2$.   We also give the minimum flux density, $S_{\rm min}$, required to detect an absorption line with $F_{\rm th}$ at $\rm S/N=5$.   Additionally, we give the expected number of background sources in the sky at $z\simeq 6$ with $S_{\rm min}$ reported by \citet{Saxena_2017} for an observing time of $100\rm\,hrs$ with the standard LOFAR configuration (see their fig. 11).  
As a quantitative example, using Table~\ref{tab:observability} (SKA1-low), for a ten background sources with $S_{203\rm\,MHz}=3.4\rm\,mJy$, on average we would expect to detect at least one \21cm absorption line with $F<0.95$ at $z=6.0\pm 0.1$ if $f_{\rm X}\leq 0.007$.    A null-detection would instead imply a lower limit of $f_{\rm X}>0.007$.  Within our model, this X-ray efficiency may be converted to an estimate of the soft X-ray band emissivity at $0.5$--$2\rm\,keV$, where $\epsilon_{\rm X,0.5-2\rm\,keV}=10^{38.3}f_{\rm X} \rm\,erg\,s^{-1}\,cMpc^{-3}$
at $z=6$.  Hence $f_{\rm X}>0.007$ corresponds to $\epsilon_{\rm X,0.5-2\rm\,keV}>10^{36.1}\rm\,erg\,s^{-1}\,cMpc^{-3}$.  Alternatively, from Table~\ref{tab:observability2} (LOFAR), the null-detection of an absorption line with $F<0.95$ at $z=6.0\pm 0.1$ in the spectra of $10$ radio bright sources with $S_{203\rm\,MHz}=18.2\rm\,mJy$ would imply a slightly weaker constraint of $f_{\rm X}>0.001$ and $\epsilon_{\rm X,0.5-2\rm\,keV}>10^{35.3}\rm\,erg\,s^{-1}\,cMpc^{-3}$.  This suggests that lower limits on the soft X-ray background emissivity at high redshift from a null-detection of the \21cm forest may complement existing constraints from upper limits on the \21cm power spectrum \citep{Greig_2020_MWA}.  We note, however, these results are highly model dependent.  If the \Lya coupling is very weak (i.e. if $f_{\alpha}\ll 1$), or there is a significant contribution to the \21cm forest absorption from unresolved small scale structure, the $f_{\rm X,max}$ values in Table~\ref{tab:observability}--\ref{tab:observability3} will translate to lower limits on $f_{\rm X}$ that are conservative.


\section{Conclusions}\label{sec:conclusion}

We have used very high resolution hydrodynamical simulations combined with a novel approach for modelling patchy reionisation to model the \21cm forest during the epoch of reionisation.  Our simulations have been performed as part of the Sherwood-Relics simulation programme (Puchwein et al. in prep).  In particular, we have considered the observability of strong ($\tau_{21}>10^{-2}$) \21cm absorbers in a late reionisation model consistent with the large \Lya forest transmission fluctuations observed at $z=5.5$ \citep{Becker_2015}, where large neutral islands of intergalactic gas persist until $z\simeq 6$ \citep{Kulkarni_2019,Keating_2020}. We also explore a wide range of assumptions for X-ray heating in the pre-reionisation intergalactic medium (IGM), and have assessed the importance of several common modelling assumptions for the predicted incidence of strong \21cm absorbers.  Our key results are summarised as follows:

\begin{itemize}
    \item In a model of late reionisation ending at $z=5.3$, for an X-ray efficiency parameter $f_{\rm X}\lesssim 0.1$ (i.e. for relatively modest X-ray pre-heating of neutral hydrogen gas, such that the gas kinetic temperature $T_{\rm K}\lesssim 10^{2}\rm\,K$) strong \21cm absorption lines with optical depths $\tau_{21}\geq 0.01$ situated in neutral islands of intergalactic gas should persist until $z=6$.  In this case, the \21cm absorbers with the largest optical depths should arise from cold, diffuse gas with overdensities $3<\Delta<10$ and kinetic temperatures $T_{\rm K}<10^{2}\rm\,K$.  A null-detection of \21cm forest absorbers at $z=6$ may therefore place a valuable lower limit on the high redshift soft X-ray background and/or the kinetic temperature of the diffuse pre-reionisation IGM in the neutral islands.    With $\sim 10$ radio-loud active galactic nuclei now known at $5.5<z<6.5$ \citep[e.g.][]{Banados_2018,Liu_2020} and the prospect of more radio-loud sources being identified in the next few years, this possibility merits further investigation.\\  

    \item By far the largest uncertainty in models of the \21cm forest is the heating of the pre-reionisation IGM by the soft X-ray background \citep[see also][]{Mack_2012}.    In the absence of strong constraints on the soft X-ray background at $z \geq 6$, proposals to use the \21cm forest to distinguish between cosmological models (where differences between competing models are small compared to the effect of X-ray heating) will likely be restricted to redshifts prior to the build-up of the soft X-ray background.  Uncertainties in the strength of the Wouthuysen-Field coupling will also be important to consider if the \Lya\ background is significantly weaker than expected from extrapolating the observed star formation rate density to $z>6$. By contrast, we find the effect of uncertain pressure/Jeans smoothing on the \21cm absorption from the diffuse IGM should remain comparatively small. \\
    
    \item Models of the \21cm forest must include the effect of gas peculiar motions on absorption line formation to accurately predict the incidence of strong absorption features \citep[see also][]{Semelin_2016}.  Ignoring redshift space distortions  reduces the incidence of the strongest \21cm forest absorbers, and results in a maximum optical depth in the \21cm forest that is up to a factor of $\sim 10$ smaller compared to a model that correctly incorporates gas peculiar velocities.\\

    \item We present model dependent estimates for the minimum redshift path length required to detect a single, strong \21cm forest absorption feature as a function of redshift and X-ray efficiency parameter, $f_{\rm X}$ within a late reionisation model that ends at redshift $z=5.3$.  At $z=6.0\pm 0.1$ for an integration time of $t_{\rm int}=1000\rm \,hrs$ per background radio source, a null-detection of \21cm forest absorbers  with $F<0.95$ at a signal-to-noise of $\rm S/N=5$ in the spectra of 10 radio sources with $S_{203\rm\,MHz}>3.4\rm\,mJy$ ($>18.2\rm\,mJy$) using SKA1-low (LOFAR) implies
    a soft X-ray background emissivity  $\epsilon_{\rm X,0.5-2\rm\,keV}>10^{36.1(35.3)}\rm\,erg\,s^{-1}\,cMpc^{-3}$.  As the soft X-ray background at high redshift is still largely unconstrained, this suggests lower limits on the X-ray emissivity from a null-detection of the \21cm forest could provide a valuable alternative constraint that complements existing and forthcoming constraints from upper limits on the \21cm power spectrum.\\ 
\end{itemize}

While the calculation we present in this work is illustrative, a more careful forward modelling of the \21cm absorption data is still required.  We have not considered how to recover absorption features from noisy data beyond the simple signal-to-noise calculation adopted here, or how an imperfect knowledge of the radio source continuum and/or radio background might impact upon the detectability of \21cm absorbers.  Uncertainties in other parameters such as the reionisation history and the \Lya background emissivity should furthermore be marginalised over to obtain a robust lower limit on the soft X-ray background.  Our simulations do not account for the absorption from unresolved mini-haloes with masses $<2.5\times 10^{7}\,M_{\odot}$, and will lack coherent regions of neutral gas on scales greater than our box size of $40h^{-1}\rm\,cMpc$.  On the other hand, even a modest amount of feedback, either in the form of photo-evaporation \citep{Park+2016,Nakatani_2020} or feedback from star formation \citep{Meiksin_2011} will substantially reduce the absorption signature from minihaloes.  These feedback effects may be particularly important during the final stages of reionisation at $z\simeq 6$, where any remaining \21cm absorption should arise from neutral islands in the diffuse IGM.  

More detailed models the \21cm forest will require either radiation-hydrodynamical simulations that encompass a formidable dynamic range, and/or multi-scale, hybrid approaches that adopt sub-grid models for unresolved absorbers and their response to feedback.  Both must furthermore cover a very large and uncertain parameter space.  Nevertheless,  we conclude that if reionisation completes at $z<6$, the prospects for using SKA1-low or possibly LOFAR to place an independent constraint on the soft X-ray background using strong absorbers in the \21cm forest are encouraging.   

\section*{Acknowledgements}

  We thank Benedetta Ciardi, Margherita Molaro, and Shikhar Mittal for helpful discussions.  We also thank an anonymous referee for their insightful comments.  The simulations used in this work were performed using the Joliot Curie supercomputer at the Tré Grand Centre de Calcul (TGCC) and the Cambridge Service for Data Driven Discovery (CSD3), part of which is operated by the University of Cambridge Research Computing on behalf of the STFC DiRAC HPC Facility (www.dirac.ac.uk).  We acknowledge the Partnership for Advanced Computing in Europe (PRACE) for awarding us time on Joliot Curie in the 16th call. The DiRAC component of CSD3 was funded by BEIS capital funding via STFC capital grants ST/P002307/1 and ST/R002452/1 and STFC operations grant ST/R00689X/1.  This work also used the DiRAC@Durham facility managed by the Institute for Computational Cosmology on behalf of the STFC DiRAC HPC Facility. The equipment was funded by BEIS capital funding via STFC capital grants ST/P002293/1 and ST/R002371/1, Durham University and STFC operations grant ST/R000832/1. DiRAC is part of the National e-Infrastructure.  This work has made use of \texttt{matplotlib} \citep{Hunter_2007}, \texttt{astropy} \citep{Robitaille_2013}, \texttt{numpy} \citep{Harris_2020} and \texttt{scipy} \citep{Virtanen_2020}.  TŠ is supported by the University of Nottingham Vice Chancellor's Scholarship for Research Excellence (EU). JSB acknowledges the support of a Royal Society University Research Fellowship.  JSB and NH are also supported by STFC consolidated grant ST/T000171/1.  MGH acknowledges support from the UKRI STFC (grant numbers ST/N000927/1 and ST/S000623/1).  We thank Volker Springel for making \textsc{P-GADGET-3} available.

\section*{Data Availability}

All data and analysis code used in this work are available from the first author on reasonable request.  An open access preprint of the manuscript will be made available at arXiv.org.



\bibliographystyle{mnras}
\bibliography{mnras_template} 




\appendix

\section{Test of the prescription for converting dense gas into collisionless particles}\label{sec:starformation}

\begin{figure}
    \begin{minipage}{\columnwidth}
 	  \centering
 	  \includegraphics[width=\linewidth]{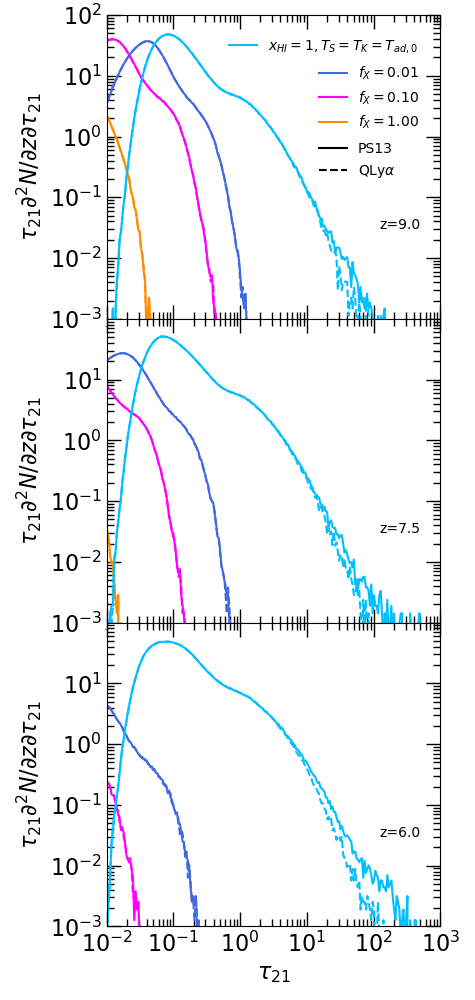}
	\end{minipage}
	\vspace{-0.5cm}
    \caption{The differential line number density distribution for \21cm forest absorption features in simulations that use two different implementations for the treatment of dense gas.  The dashed curves displays the simplified approach used in this work (the QLy$\alpha$ simulation), whereas the solid curves use the sub-grid star formation and feedback model from \citet{Puchwein_2013}  (the PS13 simulation, solid line).  The results are shown at three different redshifts: $z=9$ (top), $7.5$ (middle) and $6$ (bottom).  The X-ray efficiencies are $f_{\rm X}=0.01$ (blue curves), $0.1$ (fuchsia curves) and $1$ (orange curves).  For comparison, the cyan curves show the distribution for fully neutral, unheated gas with temperature equal to the adiabatic temperature at mean density (i.e. $x_{\rm HI}=1$ and $T_{\rm S}=T_{\rm K}=T_{\rm ad,0}$).} 
    \label{fig:starformationtest}
\end{figure}

As discussed in Section~\ref{sec:simulation}, we adopt a simplified scheme for the treatment of dense, star forming gas in the Sherwood-Relics simulations, where all gas particles with density $\Delta>1000$ and temperature $T_{\rm K}<10^5\rm\,K$ are converted to collisionless star particles \citep{Viel_2004}.  As a consequence, very dense, cold halo gas is not included in the Sherwood-Relics models.  We test whether this approximation affects our results for the \21cm forest in Fig.~\ref{fig:starformationtest}.  Here we compare two models drawn from the Sherwood simulation suite \citep{Bolton_2017} that use the same box size, mass resolution and initial conditions as the other simulations used in this work.  These two additional simulations use either the simplified scheme used in this study (QLy$\alpha$) or the star formation and energy driven winds prescription of \citet[][PS13]{Puchwein_2013}.   The only difference between these two models is the incorporation of dense, star forming gas within the PS13 simulation.   

In Fig.~\ref{fig:starformationtest}, we show the differential line number distribution obtained after applying the neutral fraction, gas kinetic and spin temperature from the patchy zr53 simulation to the native density and peculiar velocity fields from the QLy$\alpha$ and PS13 models.  As before, we consider three different X-ray efficiencies.   We observe little to no difference in the statistics of the \21cm forest computed from these two simulations. This is because the highest density gas is usually located close to ionising sources, and so is often too hot, ionised or rare to exhibit significant amounts of strong absorption in the hyperfine line.  This is further illustrated by the cyan curves in Fig.~\ref{fig:starformationtest}, where the mock \21cm forest spectra are instead computed assuming a fully neutral, isothermal gas with $T_{\rm S}=T_{\rm K}=T_{\rm ad,0}$, where $T_{\rm ad,0}=2.73\rm\, K(1+\mathnormal{z})^2/148.8$ is the gas temperature assuming adiabatic cooling at the mean density.  Small differences due to the presence of the high density gas in the PS13 simulation are now apparent in the tail of the distribution at $\tau_{21}\gtrsim 10$.  However, if we also include the adiabatic heating of the gas by compression, such that $T_{\rm ad}=T_{\rm ad,0}(1+\delta)^{2/3}$, these models become almost identical.  We conclude that the approximate treatment of dense, star forming gas we adopt in this work should not significantly change our key results.   The relative rarity of \21cm absorption from cold gas within massive haloes suggests this population will in any case be completely dominated by \21cm absorbers from the diffuse IGM and/or minihaloes during reionisation.

\section{Calculation of the X-ray and \texorpdfstring{L\lowercase{y}$\alpha$}{} specific intensities, gas kinetic temperature and ionisation state}\label{sec:IGM_heating}

In this section we describe the model for the X-ray heated IGM introduced in Section~\ref{sec:Xray_back}.  The X-ray background is primarily responsible for ionising and heating the intergalactic medium prior to reionisation.  The proper specific
intensity of the X-ray background, $J_{\rm X,\nu}$ $[\rm erg\,s^{-1}\,cm^{-2}\,Hz^{-1}\,sr^{-1}]$, is given by the solution to the cosmological radiative transfer equation
\citep{Haardt_Madau_1996,Mirocha_2014}
\begin{equation} J_{\rm X,\nu}(z) = \frac{c(1+z)^{3}}{4\pi}\int_{z}^{z_{\star}}\frac{\epsilon_{\rm X,\nu^{\prime}}(z^{\prime})}{H(z^{\prime})(1+z^{\prime})}e^{-\bar{\tau}_{\nu}(z,z^{\prime})}\,dz^{\prime}, \label{eq:Jnu} \end{equation}
\noindent
where $\epsilon_{X,\nu}$ is the comoving X-ray emissivity, $z_{\star}$ is the
redshift when X-ray emitting sources first form, and the emission
frequency, $\nu^{\prime}$ of a photon emitted at redshift $z^{\prime}$
and observed at frequency $\nu$ and redshift $z$ is
\begin{equation} \nu^{\prime}=\nu\frac{(1+z^{\prime})}{(1+z)}. \end{equation}
\noindent
The optical depth encountered by a photon observed at frequency $\nu$ is
\begin{equation} \bar{\tau}_{\nu}(z,z^{\prime})=c\sum_{i}\int_{z}^{z^{\prime}} \frac{\bar{n}_{i}(z^{\prime \prime})\sigma_{\nu^{\prime \prime},i}}{H(z^{\prime \prime})(1+z^{\prime \prime})}\,dz^{\prime \prime}, \label{eq:tau} \end{equation}
\noindent
where the sum is over the species $i=\HI,\,\HeI,\,\HeII$, and
$\sigma_{\nu,i}$ are the photo-ionisation cross-sections \citep{Verner_1996}.

The photo-ionisation rates for species $i=\HI,\,\HeI,\,\HeII$ are
\begin{align}  \Gamma_{\rm i} =~& 4\pi\int_{\nu_{i}}^{\infty}\frac{J_{\rm X,\nu}}{h_{\rm p}\nu}\sigma_{\nu,i}\,d\nu \nonumber \\
  & + 4\pi\sum_{j} \Phi_{i}(h_{\rm p}(\nu-\nu_{j}), x_{\rm e}) \int_{\nu_{j}}^{\infty}\frac{J_{\rm X,\nu}}{h_{\rm p}\nu}\sigma_{\nu,j}\,d\nu, \label{eq:PIrate} \end{align}
\noindent
where $\nu_{i}$ is the frequency of the ionisation threshold for
species $i$.  The second term in Eq.~(\ref{eq:PIrate}) arises from
secondary ionisations due to collisions with energetic
photo-electrons, where $\Phi_{\rm i}$ is the number of secondary
ionisations per primary photo-electron of energy $h_{\rm
  p}(\nu-\nu_{\rm i})$ for a free electron fraction of $x_{\rm e}$
\citep{Shullvan_Steenberg_1985}.  The corresponding photo-heating rates
are
\begin{equation} \mathscr{H}_{i} =4\pi  n_{i} \phi_{\rm heat}(h_{\rm p}(\nu-\nu_{i}), x_{\rm e})\int_{\nu_{\rm i}}^{\infty}\frac{J_{\rm X,\nu}(\nu-\nu_{i})}{\nu}\sigma_{\nu,i}\,d\nu,\end{equation}
\noindent
where $\phi_{\rm heat}$ is the fraction of the primary photo-electron
energy that contributes to the heating of the gas.  We use the
tabulated results from \citet{Furlanetto_2010} for $\Phi_{\rm
  i}$ and $\phi_{\rm heat}$.

The Compton scattering of free electrons
off X-ray background photons will also heat the IGM
\citep{Madau_1999}.  The Compton heating rate is
\begin{equation} \mathscr{H}_{\rm C}= \frac{4\pi n_{\rm e}\sigma_{\rm T}}{m_{\rm e}c^{2}}\int_{0}^{\infty}J_{\rm X,\nu}(h_{\rm p}\nu - 4k_{\rm B}T_{\rm K})\,d\nu, \end{equation}
\noindent
where $\sigma_{\rm T}=6.65\times 10^{-25}\rm\,cm^{2}$ is the Thomson
cross-section, appropriate for X-ray photons with energy $\la
100\rm\,keV$ (i.e. relativistic effects may be ignored). 

The \Lya background has two contributions: emission from stars, and
\Lya photons produced by the excitation of \HI\ atoms by X-ray
photons.  The proper \Lya specific intensity from stars,
$J_{\alpha,\star}$, requires consideration of both \Lya and higher
order Lyman series photons.  This is because the Ly$n$ photons
redshift into resonance at redshift $z$ and generate \Lya photons via
a series of radiative cascades to lower energies
\citep{Pritchard_Furlanetto_2006}, such that
\begin{equation} J_{\alpha,\star}(z)=\frac{c(1+z)^{3}}{4\pi}\sum_{n=2}^{n_{\rm max}}f_{\rm n}\int_{z}^{z_{\rm max}(n)}\frac{\epsilon_{\alpha,\nu_{\rm n}^{\prime}}(z^{\prime})}{H(z^{\prime})(1+z^{\prime})}\,dz^{\prime}, \label{eq:Jnu_alpha} \end{equation}
\noindent
where $f_{\rm n}$ is the probability of producing a \Lya photon
from a cascade from level $n$, $\nu_{\rm n}^{\prime}$ is the emission
frequency at redshift $z^{\prime}$ that corresponds to absorption by
level $n$ at redshift $z$,
\begin{equation} \nu_{\rm n}^{\prime} = \nu_{\rm LL}\left(1-\frac{1}{n^{2}}\right)\left(\frac{1+z^{\prime}}{1+z}\right), \end{equation}
\noindent
and $z_{\rm max}(n)$ is the maximum redshift from which an emitted
photon will redshift into the Ly$n$ resonance,
\begin{equation} z_{\rm max}(n) = (1+z)\frac{1-(n+1)^{-2}}{1-n^{-2}} - 1. \end{equation}
\noindent
We use the tabulated values for $f_{\rm n}$ from
\citet{Pritchard_Furlanetto_2006} and assume $n_{\rm max}=23$ \citep{Barkana_Loeb_2005}.  The
contribution from \HI\ excitation by X-ray photons is
\citep{Pritchard_Furlanetto_2007}
\begin{equation} J_{\alpha,\rm X}(z) =\frac{\lambda_{\alpha}}{4\pi H(z)} \sum_{i}\frac{\phi_{\alpha}(h_{\rm p}(\nu-\nu_{i}), x_{\rm e})}{\phi_{\rm heat}(h_{\rm p}(\nu-\nu_{i}), x_{\rm e})}\mathscr{H}_{i},  \end{equation}
\noindent
where $\phi_{\alpha}$ is the fraction of the primary photo-electron
energy that is deposited in \Lya photons
\citep{Furlanetto_2010}.

The \Lya background photons will also heat the IGM by
scattering off \HI\ atoms
\citep{Chen_2004,Chuzhoy_Shapiro_2007,Ciardi_Salvaterra_2007,Mittal_2020}, although this effect is
usually small compared to heating by X-ray photons \citep{Ciardi_2010}.  The \Lya heating rate is
\begin{equation} \mathscr{H}_{\alpha}=\frac{4\pi bH(z)}{c\lambda_{\alpha}}\left(J_{\alpha,\star,\rm c}(z)I_{\rm c}+ [J_{\alpha,\star,\rm i}(z)+J_{\alpha,\rm X}(z)]I_{\rm i}\right), \end{equation}
\noindent
where $J_{\alpha,\star,\rm c}$ is the specific intensity of continuum
($n=2$) \Lya photons, $J_{\alpha,\star,\rm i}$ is the specific intensity
of recombination photons injected at the line centre ($n>2$), and
$I_{\rm c}$, $I_{\rm i}$ are the integrals over the \Lya line
profile.  We use the approximations provided by
\citet{Furlanetto_Pritchard_2006} for $I_{\rm c}$ and $I_{\rm i}$.

Given the photo-ionisation and heating rates, the evolution of the ionisation and thermal state of the IGM at fixed gas density may be obtained by solving four coupled differential equations \citep{Bolton_Haehnelt_2007}.  We assume all gas is initially neutral and has a kinetic temperature set by adiabatic heating and cooling only,
\begin{equation} T_{\rm ad}=T_{\rm ad,0}(1+\delta)^{2/3}=2.73\rm\,K \frac{(1+z)^{2}(1+\delta)^{2/3}}{(1+z_{\rm dec})}, \label{eq:Tad} \end{equation}
where we assume the gas with overdensity $\delta$ thermally decouples from the CMB at $z_{\rm dec}=147.8$ \citep{Furlanetto_Oh_2006}.  The first three differential equations then describe the number density of ionised hydrogen, singly ionised and double ionised helium,
\begin{equation}\label{eq:DE_HII}
\frac{dn_{\rm HII}}{dt}=n_{\rm HI}(\Gamma_{\rm HI}+n_{\rm e}\Gamma_{\rm c,HI})-n_{\rm HII}n_{\rm e}\alpha_{\rm HII}, \end{equation}
\begin{align}\label{eq:DE_HeII}
    \frac{dn_{\rm HeII}}{dt} =~& n_{\rm HeI}(\Gamma_{\rm HeI}+n_{\rm e}\Gamma_{\rm c,HeI}) - n_{\rm HeII}n_{\rm e}(\alpha_{\rm HeII}+\alpha_{\rm d}) \nonumber \\
    & -\frac{dn_{\rm HeIII}}{dt},
\end{align}
\begin{equation}\label{eq:DE_HeIII}
\frac{dn_{\rm HeIII}}{dt}=n_{\rm HeII}(\Gamma_{\rm HeII}+n_{\rm e}\Gamma_{\rm c,HeII})-n_{\rm HeIII}n_{\rm e}\alpha_{\rm HeIII}, \end{equation}
\noindent
where $\alpha_{\rm i}$ and $\Gamma_{{\rm c},i}$ are, respectively, the recombination rates \citep{Verner_Ferland_1996} and collisional ionisation rates \citep{Voronov_1997} for species $i=\HI,\,\HeI,\,\HeII$, and $\alpha_{\rm d}$ is the \HeII\ dielectronic recombination coefficient \citep{Aldrovandi_1973}. The number density of neutral hydrogen,
neutral helium  and free electrons is
\begin{equation} n_{\rm HI}= n_{\rm H}-n_{\rm HII}, \end{equation}
\begin{equation} n_{\rm HeI}= \frac{Y_{\rm p}}{4(1-Y_{\rm p})}n_{\rm H}-n_{\rm HeII}-n_{\rm HeIII}, \end{equation}
\begin{equation} n_{\rm e}= n_{\rm HII}+n_{\rm HeII}+2n_{\rm HeIII}, \end{equation}
\noindent
where $Y_{\rm p}=0.24$ is the primordial helium fraction by mass
\citep{Hsyu_2020}.  The fourth differential equation describes the kinetic
temperature for gas at fixed overdensity
\begin{equation} \frac{dT_{\rm K}}{dt}=\frac{2\mu m_{\rm H}}{3k_{\rm B}\rho}[\mathscr{H}_{\rm tot}-\Lambda_{\rm tot}]+\frac{T_{\rm K}}{\mu}\frac{d\mu}{dt}-2H(t)T_{\rm K}, \label{eq:tempevol} \end{equation}
\noindent
where $\mu$ is the mean molecular weight and $\mathscr{H}_{\rm tot}$, $\Lambda_{\rm tot}$ are the total heating and cooling rates per unit volume, respectively.\footnote{Note that for X-ray heated gas we can safely neglect the missing $[2T_{\rm K}/3(1+\delta)](d \delta/dt)$ term in Eq.~(\ref{eq:tempevol}), as this is small compared to the photo-heating term for gas in the diffuse IGM.  Instead, prior to any X-ray or \Lya heating, we just assume the gas kinetic temperature follows the solution of Eq.~(\ref{eq:tempevol}) for adiabatic heating and cooling (i.e. Eq.~(\ref{eq:Tad})).  This simplification is advantageous, as a non-local calculation of the heating from adiabatic compression is significantly more complex \citep[see also the discussion of this point in][]{VillanuevaDomingo2020}.  In practice, however, we find that even if we ignore the heating from adiabatic compression and assume an initially isothermal IGM, the change to our results is negligible. This is because all our heating models have experienced appreciable X-ray and \Lya heating by $z=6$.  Finally, recall also that heating from adiabatic compression and shocks for gas with temperatures greater than the $T_{\rm K}$ predicted by Eq.~(\ref{eq:tempevol}) is already included self-consistently within our hydrodynamical simulations.}

 The total heating rate is
\begin{equation} \mathscr{H}_{\rm tot}= \mathscr{H}_{\alpha}+\mathscr{H}_{C}+\sum_{i} \mathscr{H}_{i},\end{equation}
\noindent
and the total cooling rate is
\begin{equation} \Lambda_{\rm tot}= \sum_{i}\Lambda_{{\rm c},i} + \sum_{i}\Lambda_{{\rm ex},i} + \sum_{j}\Lambda_{{\rm rec},j} +\Lambda_{\rm brem} + \Lambda_{\rm C}, \end{equation}
\noindent
where the sums are over species $i=\HI,\,\HeI,\,\HeII$ and $j=\HII,\,\HeII,\,\HeIII$.  We consider
contributions to the total cooling rate from
collisional ionisation, collisional excitation, recombination, bremsstrahlung and
inverse Compton scattering of electrons off CMB photons, respectively
\citep[cf.][]{Katz_1996}.  We use the collisional excitation cooling
rates from \citet{Cen_1992}, the inverse Compton cooling rate from
\citet{Weymann_1965} and the bremsstrahlung cooling rate from
\citet{Theuns_1998}.  The recombination and collisional ionisation
cooling rates are derived from the \citet{Verner_Ferland_1996} and
\citet{Voronov_1997} fits, respectively.

\section{Effect of peculiar velocities, pressure smoothing and reionisation redshift and on individual 21cm absorbers}\label{sec:specparam}

\begin{figure*}
    \begin{minipage}{2\columnwidth}
 	  \centering
 	  \includegraphics[width=\linewidth]{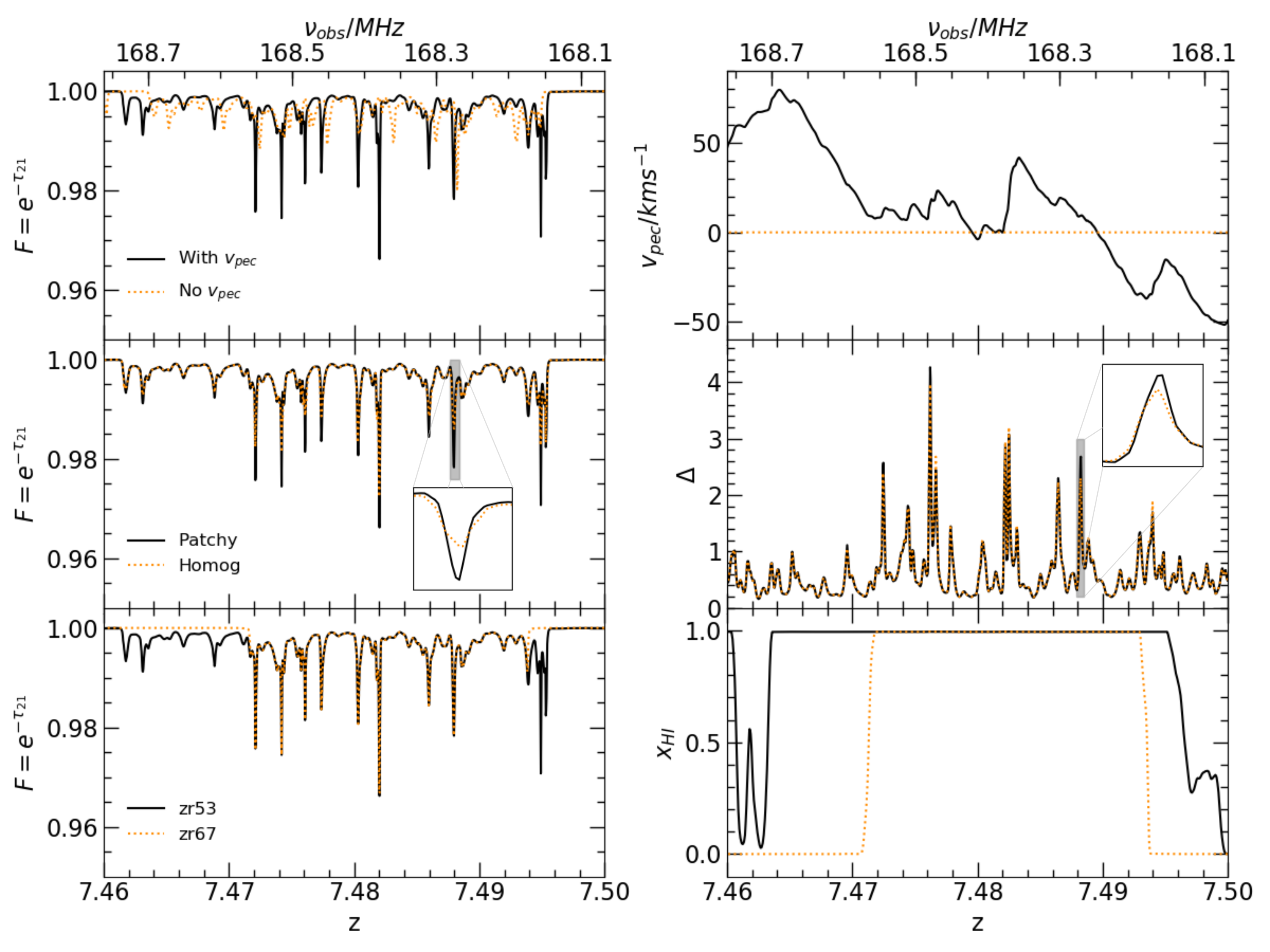}
	\end{minipage}
	\vspace{-0.3cm}
    \caption{Left: An example line of sight drawn from our mock \21cm forest spectra at $z\sim7.5$ for an X-ray efficiency of $f_{\rm X}=0.1$ and \Lya efficiency $f_{\alpha}=1$.   From top to bottom, we show the zr53 model with the solid black curve, and compare this to several model parameter variations (orange dotted curves): gas peculiar velocities set to zero (top), pressure smoothing under the assumption of homogeneous heating in the zr53-homog simulation (middle) and an earlier end to reionisation in the zr67 model (bottom). Right: The quantities responsible for the observed differences between the spectra displayed in the left column. From top to bottom, these are the gas peculiar velocity, $v_{\rm pec}$, normalised gas density $\Delta=(1+\delta)=\rho/\langle \rho \rangle$, and hydrogen neutral fraction, $x_{\rm HI}$.   In the middle panels we also present a zoomed-in view of an absorption feature (left) and the associated density peak (right), which has been broadened by pressure smoothing in the simulation with homogeneous heating.} 
    \label{fig:spectrum_compare}
\end{figure*}

In Section~\ref{sec:diffdist} and Fig.~\ref{fig:AoFdiff}, the expected differences in the incidence of strong \21cm absorbers for different modelling assumptions was discussed.  This included the effect of gas peculiar velocities, pressure (Jeans) smoothing and the timing of reionisation on the \21cm forest.  In Fig.~\ref{fig:spectrum_compare}, the impact of these modelling assumptions on individual \21cm absorbers is illustrated.

\bsp	
\label{lastpage}
\end{document}